\documentclass[12pt,onecolumn]{IEEEtran}

\usepackage{makecell}
\usepackage{array}
\usepackage{graphicx,amssymb,amsmath}
\usepackage{multicol}
\usepackage[noadjust]{cite}
\usepackage{setspace}
\usepackage{subfigure}
\usepackage{graphicx}
\usepackage{float}
\usepackage {url}
\usepackage{stfloats}
\usepackage{amsthm,pifont}
\usepackage{flushend}
\usepackage{cases,subeqnarray}
\usepackage{bm,multirow,bigstrut}
\usepackage{amsmath, amsthm, amssymb}
\usepackage{textcomp}
\usepackage{latexsym,bm}
\usepackage{booktabs}
\usepackage{xcolor}
\usepackage{mathtools}
\usepackage{dsfont}
\usepackage{extarrows}
\usepackage{epsfig}
\usepackage{epsfig}
\usepackage{algpseudocode}

\usepackage{algorithmicx,algorithm}

\usepackage{cite}
\usepackage{bm}
\usepackage{cleveref}
\usepackage{multicol}       
\usepackage{multirow}       
\usepackage{array}          
\usepackage{colortbl}
\usepackage{makecell}
\definecolor{crimson}{RGB}{192,0,0}         
\definecolor{navy}{RGB}{47,85,151}         
\usepackage{bbding}
\usepackage{graphicx}

\usepackage{booktabs}
\theoremstyle{plain}

\theoremstyle{plain}

\usepackage{amsmath}

\usepackage{threeparttable}
\usepackage{pdfsync}
\usepackage{flushend}

\IEEEoverridecommandlockouts

\begin{document}

\title{Deep Learning Based Near-Field User Localization with Beam Squint in Wideband XL-MIMO Systems}
\author{Hao Lei, Jiayi Zhang,~\IEEEmembership{Senior Member,~IEEE}, Huahua Xiao, \\ Derrick Wing Kwan Ng,~\IEEEmembership{Fellow,~IEEE}, and Bo Ai,~\IEEEmembership{Fellow,~IEEE}

\thanks{Hao Lei, Jiayi Zhang, and Bo Ai are with the State Key Laboratory of Advanced Rail Autonomous Operation, Beijing Jiaotong University, Beijing 100044, China, and also with the School of Electronics and Information Engineering, Beijing Jiaotong University, Beijing 100044, China (e-mail: haolei@bjtu.edu.cn; jiayizhang@bjtu.edu.cn; boai@bjtu.edu.cn).

H. Xiao is with ZTE Corporation, State Key Laboratory of Mobile Network and Mobile Multimedia Technology.

D. W. K. Ng is with the School of Electrical Engineering and Telecommunications, University of New South Wales, Sydney, NSW 2052, Australia (e-mail: w.k.ng@unsw.edu.au).
}
}
\maketitle

\begin{abstract}

Extremely large-scale multiple-input multiple-output (XL-MIMO) is gaining attention as a prominent technology for enabling the sixth-generation (6G) wireless networks. 
However, the vast antenna array and the huge bandwidth introduce a non-negligible beam squint effect, causing beams of different frequencies to focus at different locations.
One approach to cope with this is to employ true-time-delay lines (TTDs)-based beamforming to control the range and trajectory of near-field beam squint, known as the near-field controllable beam squint (CBS) effect.
In this paper, we investigate the user localization in near-field wideband XL-MIMO systems under the beam squint effect and spatial non-stationary properties.
Firstly, we derive the expressions for Cram\'er-Rao Bounds (CRBs) for characterizing the performance of estimating both angle and distance. 
This analysis aims to assess the potential of leveraging CBS for precise user localization.
Secondly, a user localization scheme combining CBS and beam training is proposed.
Specifically, we organize multiple subcarriers into groups, directing beams from different groups to distinct angles or distances through the CBS to obtain the estimates of users' angles and distances.
Furthermore, we design a user localization scheme based on a convolutional neural network model, namely ConvNeXt.
This scheme utilizes the inputs and outputs of the CBS-based scheme to generate high-precision estimates of angle and distance.
The numerical results derived from CRBs reveal that the inherent spatial non-stationary characteristics notably increase the CRB for angle, but have an insignificant impact on the CRB for distance estimation. In addition, the CRBs for both angle and distance decrease with increasing bandwidth and number of subcarriers.
More importantly, our proposed ConvNeXt-based user localization scheme achieves centimeter-level accuracy in localization estimates.

\end{abstract}



\IEEEpeerreviewmaketitle
\section{Introduction}

Sixth-generation (6G) wireless communications are expected to empower a variety of increasingly demanding services and applications, including Industry 5.0 and smart grid 2.0 \cite{[9]}. 
Indeed, achieving high-precision localization stands as a pivotal role in facilitating the functionality of these applications.
As such, a prominent trend involves integrating localization capabilities into future communication systems, wherein inherent communication signals are leveraged to sense the location of users or devices \cite{[10],[11],[12],[50]}.
Meanwhile, the roll-outs of extremely large-scale multiple-input multiple-output (XL-MIMO) systems, owing to their extensive deployment of antennas and enormous bandwidth, offer a promising avenue for enhancing both system capacity \cite{[16],[17],[18],[41]} and spatial resolution \cite{[13],[14],[15],[38],[40]}.
Unlike conventional far-field massive MIMO (mMIMO) systems, where the electromagnetic (EM) field is typically modeled by plane waves, in XL-MIMO systems, particularly at high frequencies, must account for near-field propagation.
In practical near-field regions, the EM field can be precisely modeled as spherical waves \cite{[16]}-\cite{[18]}, \cite{[37],[42],[43]}.
This spherical wave characteristic enables signals to be focused at specific locations through beamforming, thereby providing the potential for accurate user localization \cite{[19],[39],[3],[30]}.

In practice, as XL-MIMO systems utilize the mmWave and THz bands, employing low-cost analog hardware architecture for beamforming can lead to significant beam squint effects. 
Specifically, in XL-MIMO systems, traditional frequency-independent beamforming relying on phase shifters (PSs) can generate beams focused on specific positions, 
thereby offering enhanced spatial multiplexing gain \cite{[13]}, \cite{[15]}, \cite{[19]}.
Nevertheless, in wideband XL-MIMO systems, such beamforming based on PSs applies an identical phase shift to all subcarriers.
Consequently, the focusing area of the beams inevitably shifts in both angle and distance, causing subcarriers of different frequencies to concentrate on different areas.
This results in a considerable loss of beam focusing gain \cite{[3]}, \cite{[4]}, ultimately impacting the localization performance of wideband XL-MIMO systems.
As a remedy, several recent studies have delved into addressing beam squint effects. 
For instance, in \cite{[27]} and \cite{[28]}, true-time-delay lines (TTDs) were implemented to introduce frequency-dependent phase shifts, effectively mitigating beam squint effects and enhancing achievable rates.
On the other hand, in XL-MIMO systems, antennas positioned at different spatial locations exhibit distinct channel characteristics, described as spatial non-stationary characteristics \cite{[5]}.
Specifically, the signals received by a user equipment (UE) may only be transmitted by certain antennas of the base station (BS), which can be modeled by the visibility regions (VR).
Consequently, not all antennas of the BS are actively engaged in serving a specific UE.
This will reduce the theoretical spatial resolution of XL-MIMO systems, introducing a significant challenge for precise user localization.

\subsection{Prior Works}
Recently, several works have delved into the performance bounds for user localization in XL-MIMO systems.
For instance, the authors in \cite{[32]} derived closed-form expressions for the Cram\'er-Rao Bounds (CRBs) for angle and distance estimation in both near-field XL-MIMO radar and phased array radar modes.
Also, in \cite{[33]}, the authors evaluated the achievable estimation accuracy adopting CRBs based on vector, scalar, and overall scalar electric fields.
Similarly, the authors in \cite{[34]}, and \cite{[36]} derived the CRBs based on the received electromagnetic vector field and the near-field array response vector, respectively.
Additionally, in \cite{[35]}, the authors employed the Ziv-Zakai bound (ZZB), which incorporates prior information about unknown parameters and is not limited to unbiased estimates, to compare with the expected CRB (ECRB) and analyze the impact of noise on localization performance.
These analyses collectively demonstrate the substantial potential of XL-MIMO systems in enhancing user localization accuracy.

Existing user localization schemes for XL-MIMO systems can be divided into three categories, i.e., message passing-based, search-based, and joint channel estimation and user localization schemes.
For message passing-based schemes, the authors in \cite{[22]} constructed a probability model based on the geometric constraints of the UE location and the angles of the arrival for the subarrays.
A message passing based-algorithm was then employed to determine the UE location.
It should be noted that compared to other schemes, message passing-based schemes often have lower complexity but heavily rely on well-established probability models.
As for search-based schemes, in \cite{[23]}, the initial estimates were obtained by exploiting a maximum likelihood (ML) method, followed by iterative refinement through the trust-region algorithm to determine the final estimates of the angle and distance.
Moreover, the authors in \cite{[24]} proposed a beam-sweeping scheme for XL-MIMO systems, directing beams to cover potential location areas and estimate the user's location by maximizing received signal power.
Similarly, the authors in \cite{[25]} first constructed optimal combining matrices for each possible location to construct a pseudo-spectrum.
Subsequently, the user's location was estimated by searching for spectral peaks in the pseudo-spectrum.
Furthermore, in the downlink, the authors in \cite{[26]} designed a multiple signal classification (MUSIC) algorithm for a one-dimensional (1D) angle search and a maximum likelihood estimation (MLE) algorithm for a 1D distance search.
In the uplink, two consecutive 1D MUSIC algorithms estimated the angle and distance based on the obtained echo signal.
Interestingly, the authors in \cite{[3]} employed TTDs-based beamforming to control the degree of beam squint and proposed a low beam sweeping overhead user localization scheme, leveraging the fact that users from different locations receive maximum power at different subcarriers.
Notably, search-based solutions, while simple, have relatively higher time complexity compared to other methods.

On the other hand, for joint channel estimation and user localization schemes, the authors in \cite{[29]} initially employed variational inference to transform the channel estimation problem into an optimization problem, solved by a Newton gradient descent method. Building on this, they proposed a Gaussian fusion cooperative localization algorithm.
Similarly, in \cite{[31]}, the authors introduced a damped Newtonized orthogonal matching pursuit (DNOMP) algorithm to estimate multipath parameters and derived specific user coordinates from the spatial geometric parameters of multiple paths.
Additionally, the authors in \cite{[20]} first proposed a structured block OMP (StrBOMP) method to estimate the XL-MIMO channel and then utilized the MUSIC algorithm to obtain the high-precision user coordinates based on the estimated channel.
It is worth noting that joint channel estimation and user localization integrate the localization process into channel estimation within communication systems. This approach eliminates the need for additional signaling overhead, though it increases the complexity of the overall framework.

\vspace{-0.3cm}
\subsection{Our Contributions}
While significant efforts have been devoted to analyzing the performance bounds of user localization in XL-MIMO systems \cite{[32]}-\cite{[36]}, existing works still exhibit various shortcomings.
Firstly, the current CRB derivation overlooks the inherent spatial non-stationarity in XL-MIMO systems.
This oversight could compromise the accuracy and applicability of the bound analysis, potentially limiting its relevance for practical deployment.
Secondly, the existing CRB derivations focus primarily on the localization performance of a single carrier in the uplink of narrowband systems, neglecting the joint localization of multiple subcarriers.
Particularly in the downlink, the signal dimension received by users is typically much smaller than that received by BSs in the uplink XL-MIMO systems.
Consequently, the effectiveness of joint multi-subcarrier localization has not yet been evaluated in this context. Finally, the current CRB derivation does not consider the potential impact of beam squint in emerging wideband scenarios.
It is noteworthy that the authors in \cite{[3]} demonstrated the feasibility of exploiting different subcarriers to focus on different positions, enabling accurate estimation of user angles and distances.
Therefore, it is imperative to explore the performance lower bound of joint multi-subcarrier localization, leveraging the near-field controllable beam squint effect in spatial non-stationary channels.

Furthermore, although numerous studies have focused on designing high-precision user localization algorithms for XL-MIMO systems \cite{[3]}, \cite{[22]}-\cite{[20]}, several challenges remain unaddressed.
Firstly, most user localization schemes assume unrealistic spatial stationarity in XL-MIMO channels \cite{[3]}, \cite{[22]}-\cite{[31]}, overlooking the prevalence of spatial non-stationary characteristics in practical scenarios.
Indeed, directly applying these algorithms to spatial non-stationary channels would result in significant performance degradation and fail to satisfy practical requirements.
Secondly, several user localization schemes overlook the beam squint effect \cite{[22]}-\cite{[20]}, which can lead to significant performance degradation in future mmWave or THz communications with large bandwidths.
Although the authors in \cite{[3]} investigated low beam sweeping overhead localization algorithms based on the controllable beam squint effect, the full potential of controllable beam squint in user localization has yet to be realized.
In particular, the method proposed in \cite{[3]} did not consider spatial non-stationary characteristics.
Additionally, various factors such as noise and the accuracy of angle estimation greatly influence the localization accuracy of this scheme.
Therefore, it is necessary to further explore the application of controllable beam squint in user localization and design a high-precision user localization scheme that accounts for both beam squint effects and spatial non-stationary characteristics.

This paper investigates the user localization in a wideband XL-MIMO system, addressing both beam squint effects and spatial non-stationary properties.
Firstly, we derive expressions for the CRBs of both the angle and distance estimation. 
Moreover, we propose a low beam sweeping cost user localization scheme, which integrates the principles of the beam squint-based user localization scheme in \cite{[3]} with conventional beam training methods.
However, the aforementioned scheme does not consider the influence of spatial non-stationary characteristics, and noise significantly impacts its performance.
Additionally, the controllable beam squint-based scheme lacks explicit expressions, posing challenges for applying model-based methods to further enhance its performance.
To address these issues, we propose a deep learning (DL)-based approach to enhance the accuracy of user localization.
Specifically, we adopt the ConvNeXt model, a convolutional neural network (CNN) architecture, as detailed in \cite{[8]}. This model employs large convolutional kernels to effectively capture long-range dependencies and excels in image denoising. By leveraging these capabilities, we aim to accurately capture spatially non-stationary characteristics and minimize the impact of noise on localization accuracy.
\begin{itemize}
  \item We derive the expressions of CRBs for both angle and distance estimation based on the beam squint signal model with spatial non-stationary channels in a wideband XL-MIMO system. More importantly, we reveal the impact of spatial non-stationary effects, the number of subcarriers, and bandwidth on the CRBs, thus offering deeper insights into the factors that influence the precision of user localization.
  \item We propose a controllable beam squint-based beam training method (CBS-BT) for user localization. In this scheme, we initially direct all beams toward different angles to obtain the coarse angle estimate. Subsequently, by grouping multiple subcarriers, we steer beams from different groups towards distinct angles or distances to refine the estimates of user angles and distances.
  \item To further enhance the localization accuracy, we propose a user localization scheme utilizing the ConvNeXt in \cite{[8]} to obtain high-precision user location estimation. Specifically, the ConvNeXt-based scheme leverages large convolutional kernels and deeper network layers to alleviate the effects of spatial non-stationary and noise. This enables the extraction of angle and distance information from near-field spherical wave characteristics and controllable beam splitting effects more effectively. Moreover, with the precise angle estimation provided by the ConvNeXt, we evaluate the impact of angle estimation accuracy on the performance of the CBS-BT.
  \item The numerical results demonstrate the effectiveness of our proposed schemes. We first present the theoretical lower bound of the controllable beam squint based scheme adopting the CRBs, and analyze the factors influencing this theoretical lower bound. We have finally demonstrated that our proposed ConvNeXt based user localization scheme can achieve centimeter-level localization accuracy, even in complex spatial non-stationary mixed LoS and NLoS scenarios.
\end{itemize}

\subsection{Organization and Notation}

The rest of this paper is organized as follows. In Section \ref{System Model}, we first present the channel model and then introduce the near-field controllable beam squint.
Next, the derivation of the CRBs and two proposed user localization schemes are presented in Section \ref{User Localization}.
Then, Section \ref{Simulation Results} provides the performance analysis and numerical simulation results.
Finally, conclusions are summarized in Section \ref{Conclusions}.

{\textbf{\textit{Notation}}:}
We denote boldface lowercase letters $\rm {\bf{a}}$ and boldface uppercase letters $\rm {\bf{A}}$ as column vectors and matrices, respectively.
Conjugate, transpose, and conjugate transpose are represented as $(\cdot)^{*}$, $(\cdot)^{T}$, and $(\cdot)^{H}$, respectively.
The $ M \times N $ real-valued matrix and the $ M \times N $ complex-valued matrix are denoted by $\mathbb{R}^{ M \times N}$ and $\mathbb{C}^{ M \times N}$, respectively.
We represent the circularly symmetric complex-valued Gaussian distribution by $\mathcal{CN}(0,\sigma^2)$, where $ \sigma^2 $ is the variance.
The uniform distribution from $a$ to $b$ is denoted by $\mathcal{U} (a,b)$.
The Euclidean norm, the Hadamard product, and the $ N \times N $ identity matrix are denoted by $ \|\cdot\| $, $ \odot $, and $ {\bf{I}}_N $.
$\mathbb{E} \{ \cdot \} $ and $  {\rm{tr}}(\cdot) $ are the expectation operator the trace operator, respectively.
The $i$-th row and $j$-th column of the matrix $\rm {\bf{A}}$ is denoted by $ [{\rm {\bf{A}}}]_{i,j} $.
We denote the up rounding operation by $ \lceil \cdot \rceil $.
The absolute value of a number or the determinant of a matrix is represented as $ |\cdot| $.
We denote the real part of the matrix $\rm {\bf{A}}$ by $\Re\{\rm {\bf{A}} \}$.
The partial derivative of $x$ is denoted by $\frac{\partial }{{\partial x}}\left(  \cdot  \right)$.

\section{System Model}\label{System Model}

In this section, we present our comprehensive near-field system model. Specifically, we delve into the phenomenon of near-field beam squint and elucidate strategies for either mitigating or leveraging this effect within wideband XL-MIMO systems.

\subsection{Channel Model}

\begin{figure}
  \centering
  \includegraphics[width=2.6in]{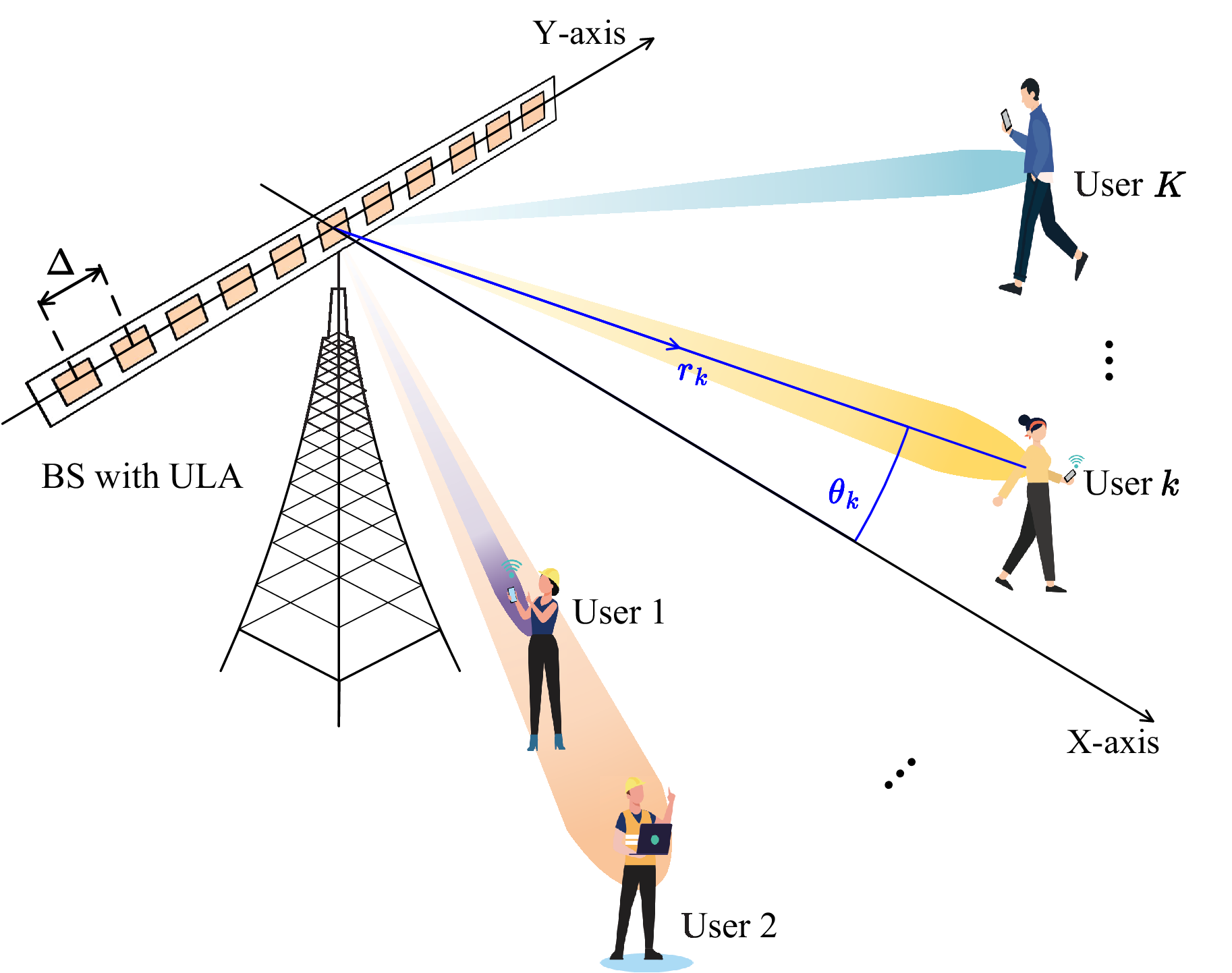}
  \caption{The wideband near-field communication scenario with a uniform linear array (ULA) at the BS in the XL-MIMO system, where UEs are single-antenna devices.}
  \label{SYSTEM-MODEL}
  \vspace{-0.4cm}
\end{figure}

With reference to Fig. 1, we consider a wideband XL-MIMO system with orthogonal frequency division multiplexing (OFDM) modulation\footnote{{{In practice, synchronization offset between the BS and UEs is crucial. However, various techniques can be exploited to achieve accurate synchronization \cite{[44]}, \cite{[46]}. Thus, this paper assumes that the synchronization offset has been resolved in line with similar studies in [20], [22]-[24], and centers on user localization with this assumption.}} }, where the base station (BS) is equipped with a $N$-element uniform linear array (ULA) to serve $K$ single-antenna users.
We denote $ \Delta = {{{\lambda _c}} \mathord{\left/  {\vphantom {{{\lambda _c}} 2}} \right.  \kern-\nulldelimiterspace} 2} $ as the antenna spacing, where $ {\lambda _c} $ denotes the central carrier wavelength.
The bandwidth and the central carrier frequency is denoted by $ B $ and $ f_c $, respectively.
By assuming there are a total of $M$ orthogonal subcarriers, the $m$-th subcarrier frequency can be represented as $ {f_m} = {f_c} + \frac{B}{M}\left( {m - 1 - \frac{{M - 1}}{2}} \right) $, $\forall m \in \left\{ {1,2, \cdots ,M} \right\}$.

Given the substantial path loss attributed to scatterers, both millimeter-wave (mmWave) or terahertz (THz) communications predominantly hinge on the line-of-sight (LoS) path \cite{[1]}, \cite{[2]}.
Consequently, our primary emphasis lies on the near-field LoS channel, with the discussions presented in this paper readily applicable to non-line-of-sight (NLoS) scenarios.
It is worth noting that unlike traditional far-field channel models, near-field XL-MIMO channel models have two key differences, i.e., spherical wave properties and spatial non-stationary properties.
To further reveal the characteristics of near-field XL-MIMO channels, a detailed discussion of these two properties is as follows.

\subsubsection{Spherical Wave Properties}

Under the spherical wavefront, the frequency-domain channel between the $n$-th BS antenna and the $k$-th UE at the $m$-th subcarrier can be expressed as
\begin{equation}
\begin{aligned}
h_{k,m}^{\left( n \right)} = \beta _{k,m}^{\left( n \right)}{e^{ - j2\pi \frac{{{f_m}}}{c}r_k^{\left( n \right)}}},
\end{aligned}
\end{equation}
where $c$ denotes the speed of light and $ {r_k^{\left( n \right)}} $ is the distance between the $n$-th BS antenna and the $k$-th UE, $\forall n \in \left\{ {1,2, \cdots ,N} \right\}$, $\forall k \in \left\{ {1,2, \cdots ,K} \right\}$.
Without loss of generality, the $x$-$y$ Cartesian coordinate of the $n$-th antenna in the BS is set as $ ( 0, n \Delta ) $.
Then, the distance $ {r_k^{\left( n \right)}} $ can be formulated as
\begin{equation}
\begin{aligned}
r_k^{\left( n \right)} &= \sqrt {{{\left( {{r_k}\cos {\theta _k}} \right)}^2} + {{\left( {{r_k}\sin {\theta _k} - n'_n\Delta } \right)}^2}} \\
&\mathop  \approx \limits^{\left( a \right)} {r_k} - n'_n\Delta \sin {\theta _k} + \frac{{{(n'_n)^2}{\Delta ^2}{{\cos }^2}{\theta _k}}}{{2{r_k}}},
\end{aligned}
\end{equation}
where $\left( {{r_k},{\theta _k}} \right)$ is the polar coordinate of the $k$-th UE, $ n'_n = \frac{{2n - N + 1}}{2}$, $\forall n \in \left\{ {1,2, \cdots ,N } \right\} $ and $(a)$ holds by the second-order Taylor expansion.
The channel fading $ \beta _{k,m}^{\left( n \right)} $ can be represented as $ \beta _{k,m}^{\left( n \right)} = \frac{c}{{4\pi {f_m}r_k^{\left( n \right)}}} $ \cite{[3]}, \cite{[4]}.
Typically, the distance $ {r_k} $, $ \forall k \in \left\{ {1,2, \cdots ,K} \right\} $, is much larger than the the array aperture of the BS ($ D = N \Delta = N \lambda_c / 2 $).
For instance, in the case of a $512$-element ULA operating at a central carrier frequency of $100$ GHz, it is highly probable that $ {r_k} > D = 512 \times 0.003 \times 0.5 = 0.768 $ meters.
In this context, we can impose an assumption that $ \beta _{k,m}^{\left( n \right)} = \beta _{k,m}^{\left( i \right)} =  \frac{c}{{4\pi {f_m}r_k}} = \beta _{k,m}$, $ \forall k, i \in \left\{ {1,2, \cdots ,K} \right\} $.
Then, the near-field LoS channel, $ {{\bf{h}}_{k,m}} \in {\mathbb{C}^{{N} \times 1 }} $, between the BS and the $k$-th UE at the $m$-th subcarrier can be formulated as
\begin{equation}
\begin{aligned}\label{3}
{{\bf{h}}_{k,m}} &= {\beta _{k,m}}\left[ {{e^{ - j2\pi \frac{{{f_m}}}{c}r_k^{\left( { - \frac{{N - 1}}{2}} \right)}}}, \cdots ,{e^{ - j2\pi \frac{{{f_m}}}{c}r_k^{\left( {\frac{{N - 1}}{2}} \right)}}}} \right]\\
&= \sqrt N {\beta _{k,m}}{\bf{a}}\left( {{r_k},{\theta _k},{f_m}} \right),
\end{aligned}
\end{equation}
where $ {\bf{a}}\left( {{r_k},{\theta _k},{f_m}} \right) = \frac{1}{{\sqrt N }}[{e^{ - j2\pi \frac{{{f_m}}}{c}r_k^{( - \frac{{N - 1}}{2})}}}, \cdots,$ ${e^{ - j2\pi \frac{{{f_m}}}{c}r_k^{(\frac{{N - 1}}{2})}}}]$ is the near-field array response vector.

The spherical wave properties introduce new opportunities for precise user localization within XL-MIMO systems, compared with their plane wave counterparts.
Specifically, electromagnetic waves in near-field XL-MIMO systems exhibit variations in both phase and amplitude across the antenna array in the received signal.
In this context, the array response vector employed in near-field channel modeling can capture both angular and distance information from the UE to the BS.
In contrast, traditional far-field channel models rely on the assumption of plane wavefront, with array response vectors containing solely angle information.
Therefore, exploiting channel modeling based on spherical waves to describe the electromagnetic characteristics, including distance and angle characteristics, significantly improves the spatial resolution of practical near-field XL-MIMO systems, thereby bringing new opportunities for improving user localization performance.

\subsubsection{Spatial Non-stationary Properties}

In conventional far-field mMIMO systems \cite{[16]}, \cite{[17]}, \cite{[18]}, the antenna array aperture of the BS tends to be relatively modest, resulting in spatially stationary channels where the channel coefficients observed by one random UE across different antennas at the BS are evenly distributed.
Conversely, near-field XL-MIMO systems feature physically larger array apertures, giving rise to spatially non-stationary channel characteristics.
In this scenario, the signal received by a UE may originate solely from specific antennas of the BS.
This distinct visible area of the array from the perspective of a UE is defined as the visibility region (VR) of that UE \cite{[5]}.
Considering the spatial non-stationary properties of near-field XL-MIMO channels, the channel in \eqref{3} can be reformulated as
\begin{equation}
\begin{aligned}\label{4}
{{\bf{h}}_{k,m}} = \sqrt N {\beta _{k,m}}{\bf{a}}\left( {{r_k},{\theta _k},{f_m}} \right)  \odot {\bf{b}}\left( {\bf{\Theta }}_k \right),
\end{aligned}
\end{equation}
where $ {\bf{b}}\left( {\bf{\Theta }}_k \right) \in {\mathbb{C}^{{N} \times 1 }} $ indicates the antennas of the BS that are within the line of sight of the $k$-th UE and $ {\bf{\Theta }}_k $ collects the indices of antennas that are visible to the $k$-th UE, which can be expressed as
\begin{equation}
\begin{aligned}\label{5}
{\left[ {{\bf{b}}\left( {{\Theta _k}} \right)} \right]_n} = \left\{ {\begin{array}{*{20}{c}}
{1,\quad n \in {\bf{\Theta }}_k}\\
{0,\quad n \notin {\bf{\Theta }}_k}
\end{array}} \right. .
\end{aligned}
\end{equation}
According to the channel measurement in \cite{[5]}, it is evident that while the overall channel exhibits spatial non-stationarity, sub-channels corresponding to specific sections of the array (referred to as sub-arrays) can be regarded as spatially stationary.
To facilitate the analysis in the sequel, we partition the ULA into $ N_s $ smaller non-overlapped sub-arrays, assuming that $ N/N_s $ is an integer without loss of generality.
Introducing the set $ {{\bf{\Xi }}_k} $ to index the sub-arrays that can establish a LoS to the $k$-th UE, we have $ {{\bf{\Xi }}_k} = \left\{ {{n_{k,1}},\cdots ,{n_{k,{S_k}}}} \right\} $, where $ {n_{k,i}} $ is the index of the sub-array with $ 1 \leq {n_{k,i}} \leq N_s $ and $ S_k $ denotes the total number of sub-arrays having a LoS to the $k$-th UE with $1 \leq  S_k  \leq N_s $.
Then, the vector $ {\bf{b}}\left( {\bf{\Theta }}_k \right) $ can be further expressed as
\begin{equation}
\begin{aligned}
{\left[ {{\bf{b}}\left( {{\bf{\Theta }}_k} \right)} \right]_n} = \left\{ {\begin{array}{*{20}{l}}
{1,  \quad\quad {\rm{if}}\left\lceil {\frac{{n{N_s}}}{N}} \right\rceil  \in {{\bf{\Xi }}_k}},\\
{0,  \quad\quad {\rm{else}}}.
\end{array}} \right.
\end{aligned}
\end{equation}

The spatial non-stationary properties introduce new challenges to accurate user localization.
Specifically, not all BS antennas serve a specific user.
This leads to a certain loss of theoretical resolution improvement caused by the large aperture array of the BS in practical XL-MIMO systems, which in turn leads to a decrease in localization performance.
Therefore, it is necessary to design a user localization scheme that can alleviate the impact of spatial non-stationary characteristics.

\subsection{Near-field Controllable Beam Squint}

\begin{figure}
\centering
\subfigure[The PSs based beamforming architecture.] {\includegraphics[height=3.5cm,width=4.35cm]{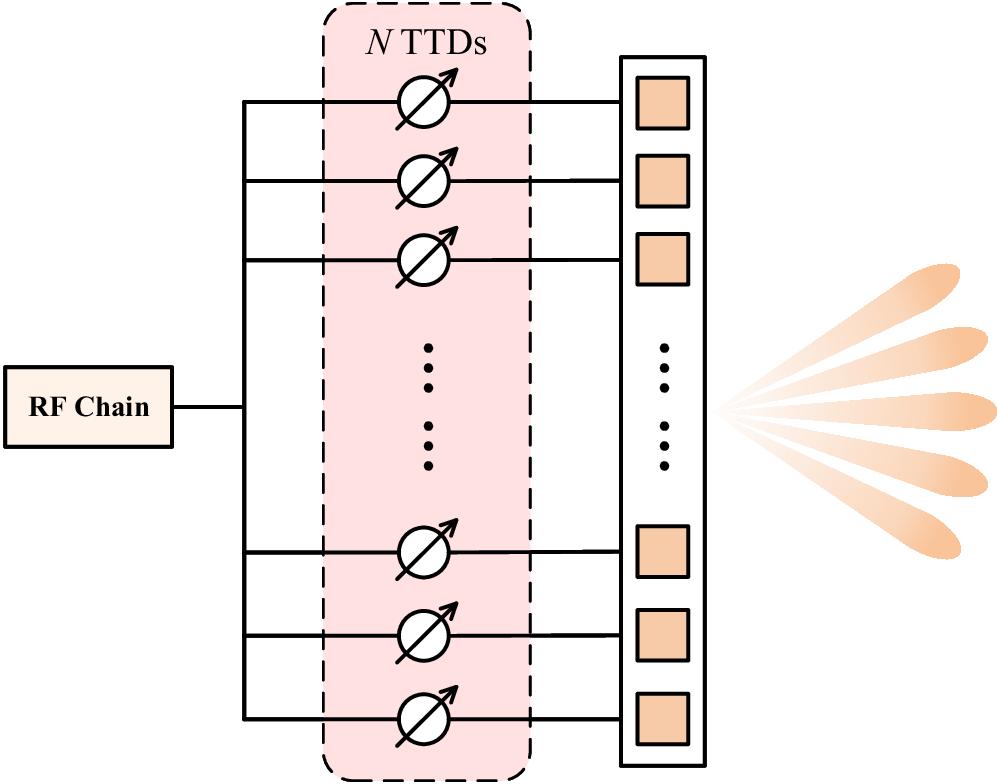}}
\subfigure[The near-field beam squint effect caused by the PSs based beamforming.] {\includegraphics[height=3.5cm,width=4.35cm]{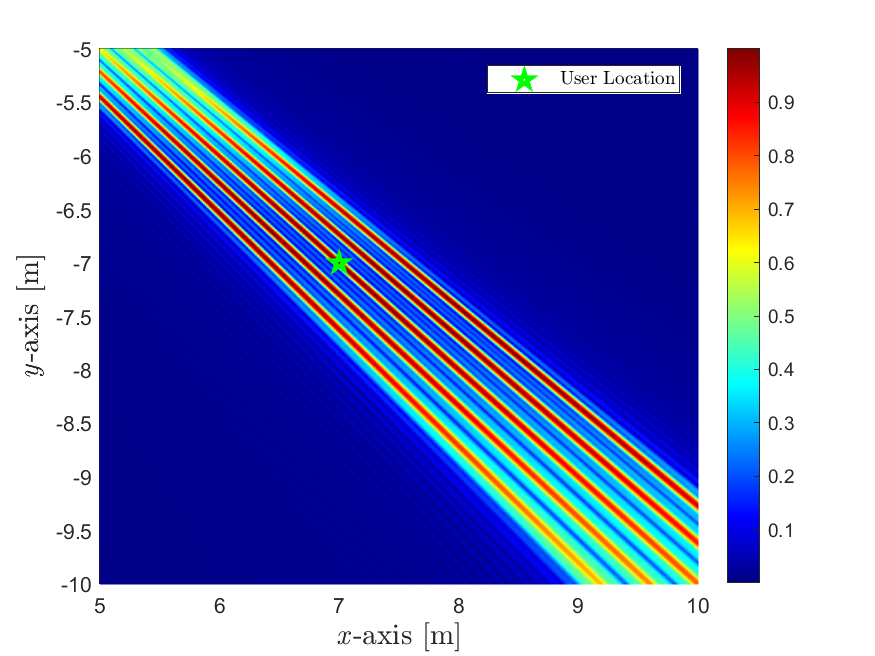}}
\subfigure[The TTD lines based beamforming architecture.] {\includegraphics[height=3.5cm,width=4.35cm]{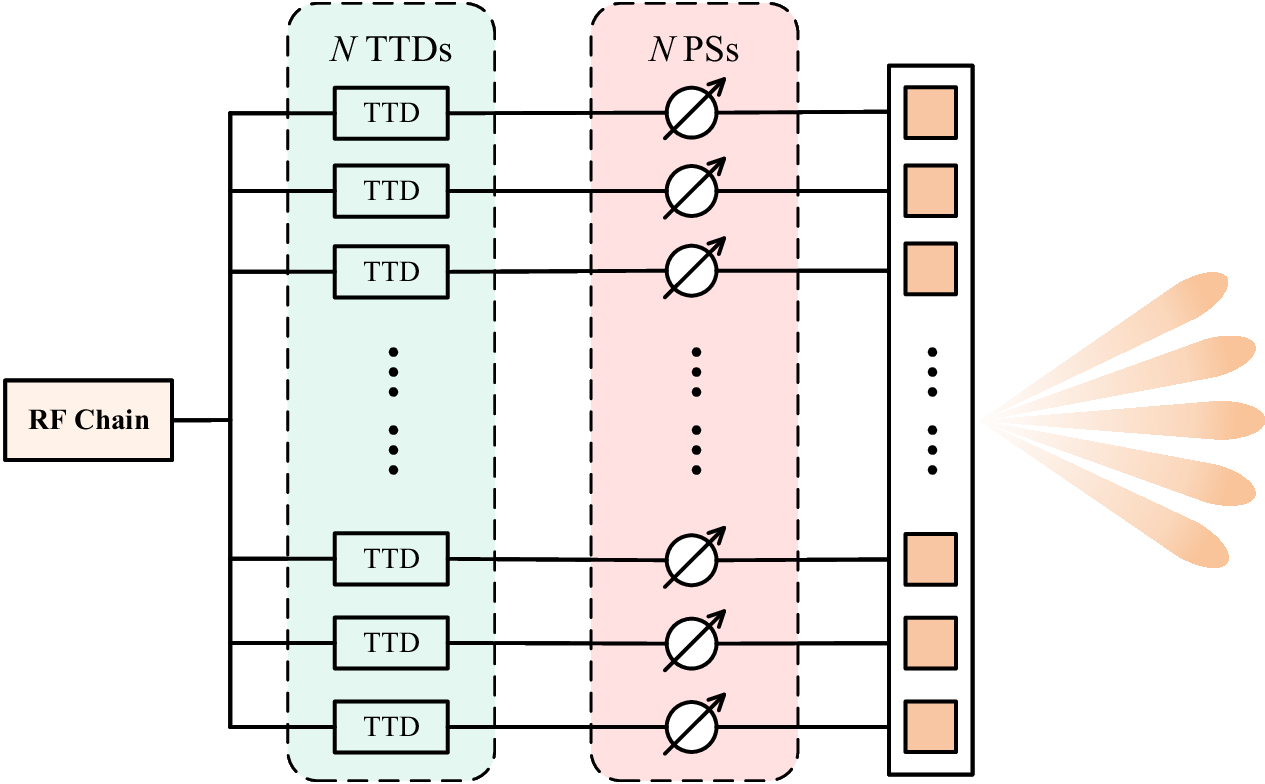}}
\subfigure[The near-field beams with the TTD lines based beamforming.] {\includegraphics[height=3.5cm,width=4.35cm]{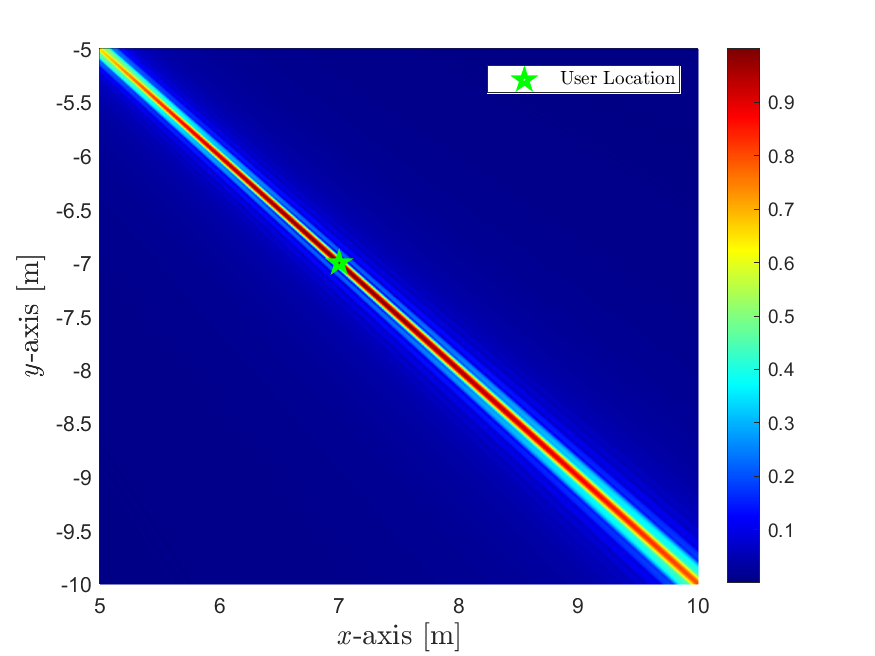}}
\caption{Illustration of the near-field beamformeing architectures and their corresponding normalized beamforming gains. The sub-figure (a) and (c) depict beamforming architectures based on PSs and TTDs, respectively. Meanwhile, the sub-figure (b) and (d) depict the normalized beamforming gains corresponding to the sub-figure (a) and (c), respectively. In sub-figure (b) and (d), blue represents the minimum gain, red represents the maximum gain, and other colors transition between these two extremes.
These representations are depicted considering a bandwidth of $B=6$ GHz, a central carrier frequency of $f_c=100$ GHz, the utilization of $M=5$ subcarriers, and an array comprising $N=512$ antennas. The location of the UE is set as $ (7, -7) $.
}
\label{architecture}
\vspace{-0.7cm}
\end{figure}

In the mmWave or THz bands, BSs commonly employ frequency-independent phase-shifters (PSs)-based beamforming to serve UEs, as shown in Fig \ref{architecture} (a).
Particularly in near-field scenarios, the BS can generate a focused beam aligned with the position ($r, \theta$) of the UE in polar coordinates, significantly enhancing spatial multiplexing capabilities.
In this context, the optimal beamfocusing vector produced by frequency-independent PSs\footnote{     {{    The implementation of ideal continuous phase shifters is quite challenging, so quantized phase shifters with limited resolution are commonly used in practical systems. However, as XL-MIMO technology is still in the initial exploration stage of user localization, this paper adopts the same continuous phase shifter model as in \cite{[3]}, \cite{[4],[27],[28]} to investigate the performance lower bounds of joint multi-subcarrier localization, leveraging the near-field controllable beam squint effect in spatially non-stationary channels. It is important to highlight that quantized phase shifters with limited resolution will be a primary focus of future work.}}} for UE $k$ can be denoted as ${\bf{w}}_k = {\bf{a}}\left( {r_k,\theta_k,f_1} \right)$, where $ {\bf{a}}\left( {{r_k},{\theta _k},{f_m}} \right) $ is the near-field array response vector at the $1$-st subcarrier.

It is worth noting that the beamfocusing vector ${\bf{w}}_k$ would generate a fixed phase shift for beams of different frequencies throughout the entire frequency band.
However, the wideband channel \eqref{4} is subject to frequency dependency.
Consequently, a mismatch arises between the aforementioned beamfocusing vector and the practical channel.
Specifically, the beamfocusing vector ${\bf{w}}_k = {\bf{a}}\left( {r_k,\theta_k,f_1} \right)$ can only focus the beam with frequency $f_1$ on the position $($$r_k, \theta_k $$)$, while beams with other frequencies would be focused on different, non-ideal positions $(r, \theta) \neq (r_k, \theta_k)$, as shwon in Fig \ref{architecture} (b).
This phenomenon is commonly known as beam split \cite{[27]}, \cite{[28]}, or beam squint effect \cite{[3]}, and in this paper, we adopt the latter terminology.

Recently, several works have focused on the beam squint effect in near-field XL-MIMO systems \cite{[3]}, \cite{[4]}, \cite{[27]}, \cite{[28]}.
One viable solution to address this issue involves the utilization of the true-time-delay (TTD) lines based beamforming instead of the PSs based beamforming.
By leveraging this approach, frequency-dependent beams can be meticulously generated to align with the characteristics of frequency-dependent channels.
Consequently, this method offers an effective means to mitigate beam squint effects in near-field XL-MIMO systems.
As shown in Fig. \ref{architecture} (c), we assume that the BS is equipped with $N$ TTD lines, where each TTD line is cascaded with a phase shifter.
Then, the TTD lines based beamfocusing vector $ {{{\bf{w}}_{k,m}}} \in {\mathbb{C}^{{N} \times 1 }} $ for UE $k$ at the $m$-th subcarrier can be reformulated as
\begin{equation}\label{7}
\begin{aligned}
{\left[ {{{\bf{w}}_{k,m}}} \right]_n} = \frac{1}{{\sqrt N }}{e^{ - j2\pi {\phi _n}}}{e^{ - j2\pi {{\tilde f}_m}{t_n}}},
\end{aligned}
\end{equation}
where $ {\phi _n} $ is the phase shift of the $n$-th PS, $ t_n $ is the time delay of the $n$-th TTD line, and $ {{\tilde f}_m}$ is a frequency-dependent term.

\begin{figure}
  \centering
  \includegraphics[width=2.6in]{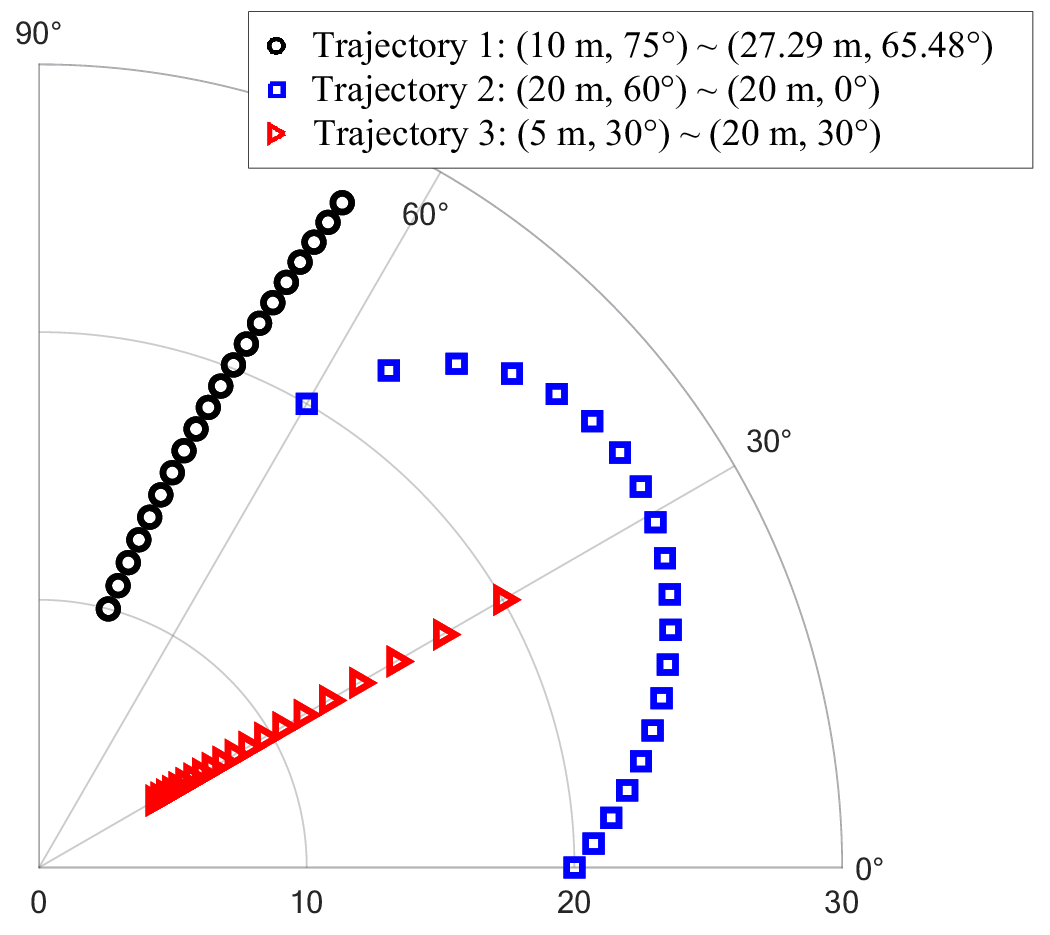}
  \caption{The near-field beam trajectories based on the controllable beam squint. Parameters for these trajectories include a a bandwidth of $B=6$ GHz, a central carrier frequency of $f_c=100$ GHz, the utilization of $M=20$ subcarriers, and an array comprising $N=512$ antennas. Taking trajectory $2$ as an example, when we set $ {r_s} = {r_e} = 20$ $\text{m}$, $ {\theta _s} = 60^\circ $, and $ {\theta _e} = 0^\circ$ in \eqref{7}, the $1$-st and the $20$-th subcarrier beams would be directed to $(20$ $\text{m}, 60^\circ) $ and $(20$ $\text{m}, 0^\circ) $, respectively. In this scenario, the focus of the $m$-th subcarrier beam can be determined by \eqref{8} and \eqref{9}. For instance, the focus of the $5$-th and $16$-th subcarrier beams is $(26.21$ $\text{m}, 42.45^\circ) $ and $(22.83$ $\text{m}, 10.01^\circ) $, respectively. These focal points corresponding to $M$ subcarriers can be connected to form a trajectory, covering a specific angle or distance range.}
  \label{T}
\vspace{-0.6cm}
\end{figure}

Note that TDD lines-based beamforming significantly mitigates the near-field beam squint effect, as shown in Fig. \ref{architecture} (d).
Moreover, it empowers the BS to precisely regulate the focusing position of both the lowest and highest frequency subcarriers.
This capability enables different subcarriers to concentrate at distinct and desired locations, as shwon in Fig. \ref{T}.
To elaborate further, in polar coordinates, when we set $ {\phi _n} = \frac{{{f_1}r_s^{\left( n \right)}}}{c} $, $ {{\tilde f}_m} = {f_m} - {f_1}$, and $ {t_n} = \frac{{{f_M}r_e^{\left( n \right)}}}{{cB}} - \frac{{{\phi _n}}}{B} $, the $1$-st subcarrier aligns its focus on $(r_s, \theta_s)$, while the $M$-th subcarrier concentrates on $(r_e, \theta_e)$.
Within this framework, the points with the maximum beamforming gain for each subcarrier can be connected into a curved trajectory, as illustrated in Fig. \ref{T}.
More specifically, the polar coordinates $\left( {{{\bar r}_m},{{\bar \theta }_m}} \right)$ of the point with the maximum beamforming gain for the $m$-th subcarrier satisfy
\begin{equation}
\begin{aligned}\label{8}
\!\!\!\!\!\!\!\!\!\!\!\!\!\!  \sin {{\bar \theta }_m} = \frac{{\left( {B - {{\tilde f}_m}} \right){f_1}}}{{B{f_m}}}\sin {\theta _s} + \frac{{\left( {B + {f_1}} \right){{\tilde f}_m}}}{{B{f_m}}}\sin {\theta _e},
\end{aligned}
\end{equation}
\begin{equation}
\begin{aligned}\label{9}
\frac{1}{{{{\bar r}_m}}} = \frac{1}{{{r_s}}}\frac{{\left( {B - {{\tilde f}_m}} \right){f_1}}}{{B{f_m}}}\frac{{{{\cos }^2}{\theta _s}}}{{{{\cos }^2}{{\bar \theta }_m}}} + \frac{1}{{{r_e}}}\frac{{\left( {B + {f_1}} \right){{\tilde f}_m}}}{{B{f_m}}}\frac{{{{\cos }^2}{\theta _e}}}{{{{\cos }^2}{{\bar \theta }_m}}}.
\end{aligned}
\end{equation}
Fig. \ref{T} illustrates several examples of near-field beam trajectories under the PSs-based and TTD lines-based beamforming.
Firstly, assuming the UE is positioned at $(10$ $\text{m}, 75^\circ)$, employing the PSs-based beamforming directs the first subcarrier beam towards $(10$ $\text{m}, 75^\circ)$.
At this juncture, it is noted that the point with the highest beamforming gain for the $M$-th subcarrier manifests at $(27.29$ $\text{m}, 65.48^\circ)$.
Evidently, the near-field beam squint effect cannot be disregarded.
Subsequently, under the TTD lines-based beamforming, trajectory 2 (trajectory 3) is obtained by setting $ {r_s} = {r_e} = 20$ $\text{m}$, $ {\theta _s} = 60^\circ $, and $ {\theta _e} = 0^\circ$ ($ {r_s} = 5$ $\text{m}$, $ {r_e} = 20$ $\text{m}$, and $ {\theta _s} = {\theta _e} = 60^\circ$) in \eqref{7}.
Observing the trajectories, trajectory 2 encompasses a notably wide range of angles, whereas trajectory 3 directs subcarrier beams of various frequencies to positions sharing the same angle but differing distances.
Leveraging this attribute, the authors in \cite{[3]} utilizes the phenomenon that UEs at distinct locations will receive maximum power at diverse subcarriers to design a beam sweeping method applicable for user localization.
In the following sections, we will further explore the potential of beam squint effects in user localization.

\vspace{-0.3cm}
\section{User Localization}\label{User Localization}

During the downlink data transmission, the received signal $ {y_{k,m}} \in {\mathbb{C}^{{1} \times 1 }} $ of the $k$-th UE on the $m$-th subcarrier can be given by
\begin{equation}
\begin{aligned}\label{10}
{y_{k,m}} = {\bf{h}}_{k,m}^H{{\bf{w}}_{k,m}}{s_{k,m}} + {n_{k,m}},
\end{aligned}
\end{equation}
where $ {{\bf{w}}_{k,m}} \in {\mathbb{C}^{{N} \times 1 }}  $ is the TTD lines based beamforming vector for the $k$-th UE on the $m$-th subcarrier, $ {s_{k,m}} \in {\mathbb{C}^{{1} \times 1 }}$ denotes the data for the $k$-th UE transmitted by the $m$-th subcarrier, $ {n_{k,m}} \sim \mathcal{CN} \left( {0,{\sigma ^2}} \right) $ denotes the additive Gaussian white noise, and $ \sigma ^2 $ denotes the noise power at the receiver.
Then, the received signal $ {y_{k,m}} \in {\mathbb{C}^{{1} \times 1 }} $ at each subcarrier can be collected and stacked into a vector, $ {{\bf{y}}_k} \in {\mathbb{C}^{{M} \times 1 }}  $, as
\begin{equation}
\begin{aligned}\label{11}
{{\bf{y}}_k} = \left[ {\begin{array}{*{20}{c}}
{{\bf{h}}_{k,1}^H{{\bf{w}}_{k,1}}{s_{k,1}}}\\
 \vdots \\
{{\bf{h}}_{k,M}^H{{\bf{w}}_{k,M}}{s_{k,M}}}
\end{array}} \right] + \left[ {\begin{array}{*{20}{c}}
{{n_{k,1}}}\\
 \vdots \\
{{n_{k,M}}}
\end{array}} \right]
 = {{\bf{u}}_k} + {{\bf{n}}_k},
\end{aligned}
\end{equation}
where $ {{\bf{u}}_k} \in {\mathbb{C}^{{M} \times 1 }} $ and ${{\bf{n}}_k} \in {\mathbb{C}^{{M} \times 1 }} $.

\vspace{-0.3cm}
\subsection{Cram\'er-Rao Bound}

Let $ { {\bm{\rho }} _k} = {\left[ {{r_k},{\theta _k}} \right]^T} $ that includes the unknown parameters in \eqref{11} and $ {{\tilde{\bm{ \rho }}}_k} $ is the unbiased estimator at the BS of $ { {\bm{\rho }} _k} $.
Then, the lower bound of the mean-squared error of the $i$-th parameter $\tilde \rho _{k,i} $ satisfies
$ {\rm{MSE}}\left\{ {{{\tilde \rho }_{k,i}}} \right\}  = {\left( {{\rho _{k,i}} - {{\tilde \rho }_{k,i}}} \right)^2} \ge {\left[ {{\rm{\bf  CRB}}\left( {{{\bm{\rho }}_k}} \right)} \right]_{i,i}} $ \cite{[7]},
where $ \left[ {{\rm{\bf  CRB}}\left( {{{\bm{\rho }}_k}} \right)} \right] $ denotes the inverse of the Fisher Information Matrix (FIM) $ \left[ {{\bf{CRB}}\left( {{{\bf{\rho }}_k}} \right)} \right] = {\left[ {{\bf{FIM}}\left( {{{\bf{\rho }}_k}} \right)} \right]^{ - 1}} $.
Since we consider the additive Gaussian white noise model, the FIM is given by \cite{[7]}
\begin{equation}
\begin{aligned}\label{13}
{\left[ {{\bf{FIM}}\left( {{{\bf{\rho }}_k}} \right)} \right]_{i,j}} = {\rm{tr}}\left[ {{\bf{R}}_k^{ - 1}\frac{{\partial {{\bf{R}}_k}}}{{\partial {\rho _{k,i}}}}{\bf{R}}_k^{ - 1}\frac{{\partial {{\bf{R}}_k}}}{{\partial {\rho _{k,i}}}}} \right] \\+ 2\Re \left[ {\frac{{\partial {\bf{u}}_k^H}}{{\partial {\rho _{k,i}}}}{\bf{R}}_k^{ - 1}\frac{{\partial {\bf{u}}_k^H}}{{\partial {\rho _{k,j}}}}} \right],
\end{aligned}
\end{equation}
where $ {\bf{R}}_k $ is the covariance matrix of the vector $ {{\bf{y}}_k} $.
Since the noise term $ {{\bf{n}}_k} \sim \mathcal{CN} \left( {0,{\sigma ^2}{\bf I}_N } \right)  $, we have $ {\bf{R}}_k = {\sigma ^2}{\bf I}_N  $.
Then, the FIM in \eqref{13} can be represented as
\begin{equation}
\begin{aligned}
{\bf{FIM}}\left( {{{\bf{\rho }}_k}} \right)\!\! =\!\! \frac{2}{{{\sigma ^2}}}\Re \left\{ {\left( {\frac{{\partial {\bf{u}}_k^H}}{{\partial {{\bf{\rho }}_k}}}} \right)\frac{{\partial {{\bf{u}}_k}}}{{\partial {{\bf{\rho }}_k}}}} \right\}
 \!\!=\!\! \frac{2}{{{\sigma ^2}}}\left[ {\begin{array}{*{20}{c}}
{{\vartheta _{k,rr}}}&{{\vartheta _{k,r\theta }}}\\
{{\vartheta _{k,\theta r}}}&{{\vartheta _{k,\theta \theta }}}
\end{array}} \right],
\end{aligned}
\end{equation}
where $ {\vartheta _{k,rr}} = \frac{{\partial {\bf{u}}_k^H}}{{\partial {r_k}}}\frac{{\partial {{\bf{u}}_k}}}{{\partial {r_k}}} $, $ {\vartheta _{k,r\theta }} = {\vartheta _{k,\theta r}} = \Re \left\{ {\frac{{\partial {\bf{u}}_k^H}}{{\partial {r_k}}}\frac{{\partial {{\bf{u}}_k}}}{{\partial {\theta _k}}}} \right\} $, and $ {\vartheta _{k,\theta \theta }} = \frac{{\partial {\bf{u}}_k^H}}{{\partial {\theta _k}}}\frac{{\partial {{\bf{u}}_k}}}{{\partial {\theta _k}}} $.
Therefore, the CRBs of the distance $ r_k $ and the angle $ \theta_k $ are given by
\begin{equation}
\begin{aligned}
{\rm{CR}}{{\rm{B}}_{{r_k}}} = \frac{{{\sigma ^2}}}{2}\frac{{{\vartheta _{k,\theta \theta }}}}{{{\vartheta _{k,rr}}{\vartheta _{k,\theta \theta }} - {\vartheta _{k,r\theta }}{\vartheta _{k,\theta r}}}},
\end{aligned}
\end{equation}
\begin{equation}
\begin{aligned}
{\rm{CR}}{{\rm{B}}_{{\theta _k}}} = \frac{{{\sigma ^2}}}{2}\frac{{{\vartheta _{k,rr}}}}{{{\vartheta _{k,rr}}{\vartheta _{k,\theta \theta }} - {\vartheta _{k,r\theta }}{\vartheta _{k,\theta r}}}},
\end{aligned}
\end{equation}
where the calculations of $ {\vartheta _{k,rr}} $, $ {\vartheta _{k,r\theta }} $, and $ {\vartheta _{k,\theta \theta }}  $ are given in Appendix.
It is evident that noise has a substantial impact on both the CRB of angle and distance.
Additionally, as demonstrated by \eqref{a4} and \eqref{a7}, increasing the number of subcarriers leads to a decrease in the root CRBs for both angle and distance.
Furthermore, spatial non-stationary characteristics and bandwidth also influence the CRB, which will be further elucidated through numerical results in Section \ref{Simulation Results}.

\vspace{-0.3cm}
\subsection{Controllable Beam Squint based Beam Training}

\begin{figure}
\centering
\subfigure[{{The normalized signal power of the angle estimation stage.}}] {\includegraphics[width=2.6in]{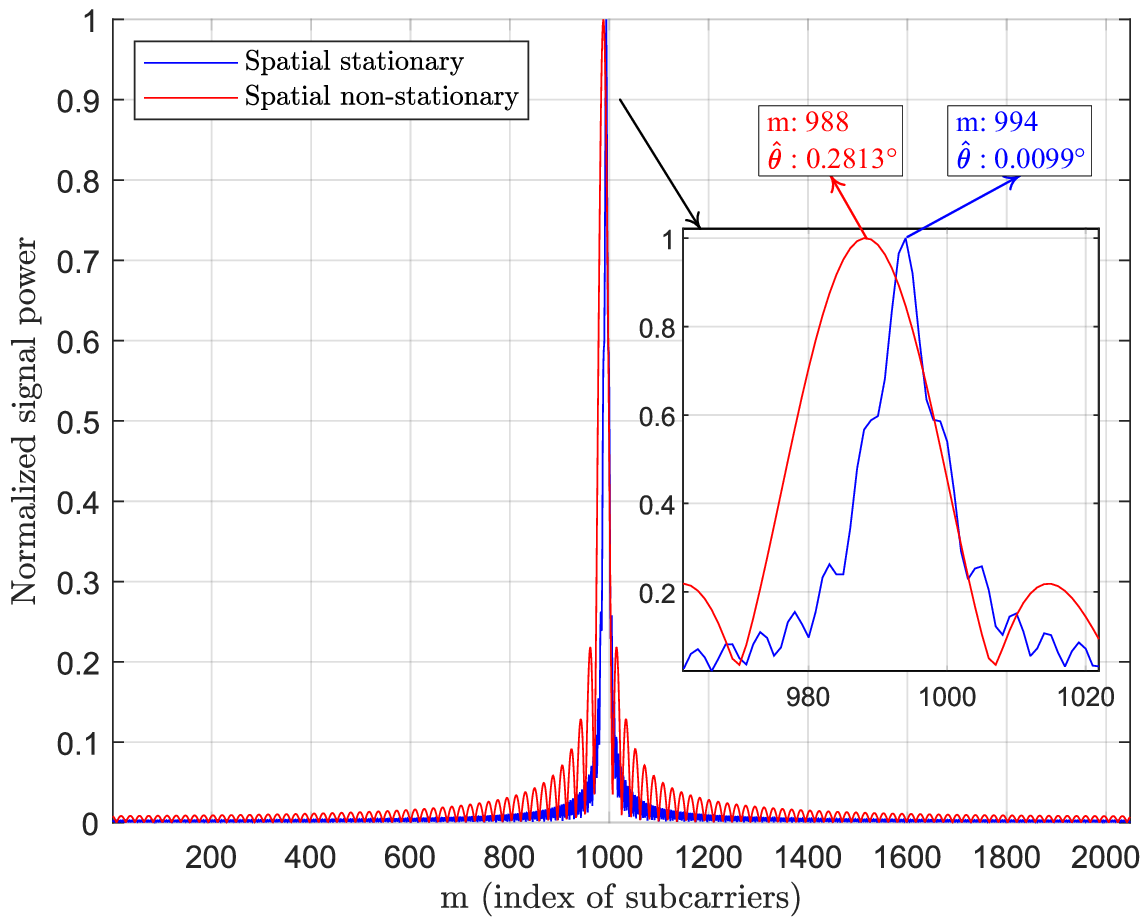}}
\subfigure[{{The normalized signal power of the distance estimation stage.}}] {\includegraphics[width=2.6in]{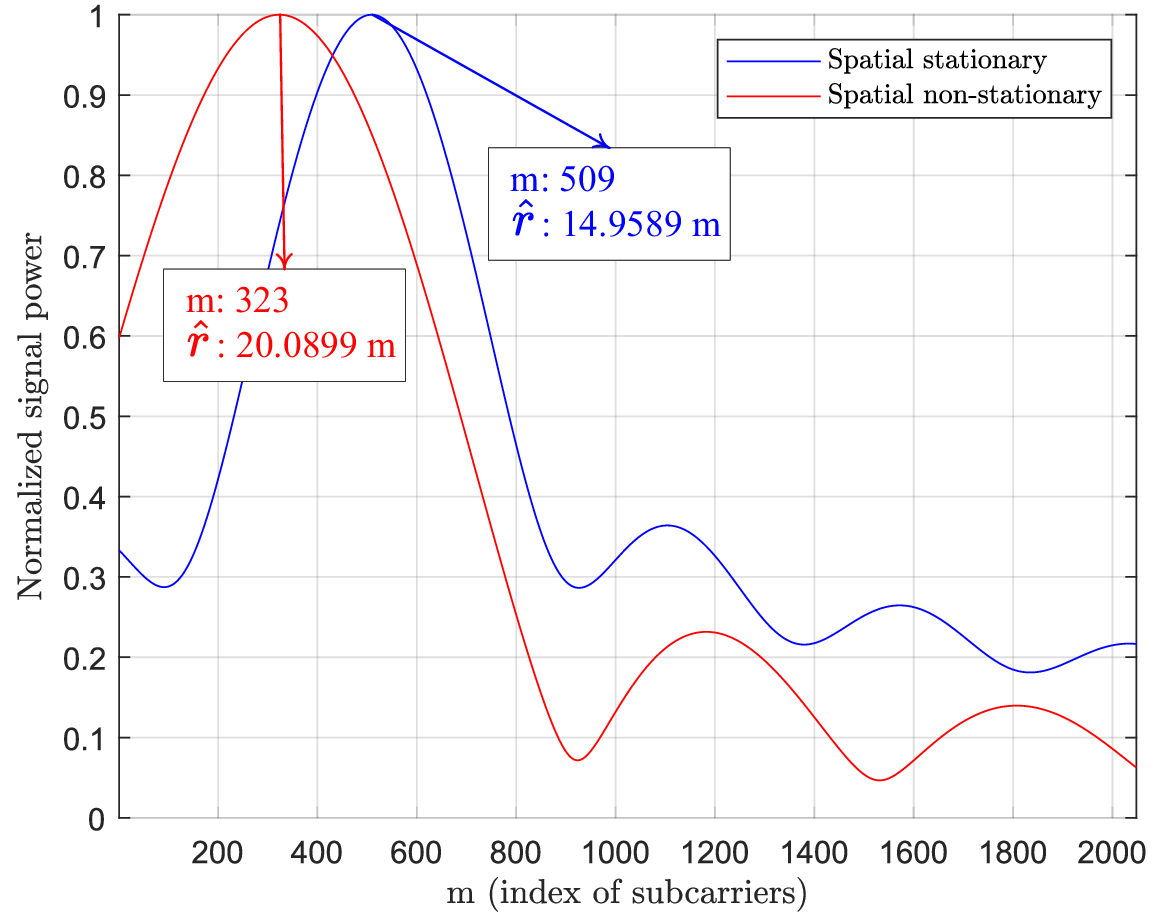}}
\caption{{{The normalized signal power of the angle estimation stage and the distance estimation stage of the scheme in \cite{[3]} under both spatial stationary channel and spatial non-stationary channel conditions, where the UE is located at $(15$ ${\rm m}, 0^\circ)$, $ M =2048 $, $ f_c = 100 $ GHz, $ N =512 $, and $ B =6 $ GHz. We assume that there is no noise for illustration.}} 
}
\label{power}
\vspace{-0.6cm}
\end{figure}

Interestingly, the authors in \cite{[3]} designed a two-stage user localization scheme with low beam sweeping overhead, capitalizing on near-field controllable beam squint as discussed in Section \ref{System Model}-B.
Assume that the locations $\left( {{r_k},{\theta _k}} \right),\forall k \in \left\{ {1,2, \cdots ,K} \right\}$, are limited by $ {r_{\min }} \le {r_k} \le {r_{\max }}$ and ${\theta _{\min }} \le {\theta _k} \le {\theta _{\max }} $.
Then, in the first stage, through adjustments of PSs and TTDs, the BS steers the first and $M$-th subcarriers towards $(r_{\text{min}}, \theta_s)$ and $(r_{\text{min}}, \theta_e)$, akin to trajectory 2 depicted in Fig. \ref{T}.
Specifically, the values of PSs and TTDs are set as
$ r_s^{\left( n \right)} = {r_{\min }} - n'_n\Delta \sin {\theta _{\min }} + \frac{{{{\left( {n'_n} \right)}^2}{\Delta ^2}{{\cos }^2}{\theta _{\min }}}}{{2{r_{\min }}}} $ and $ r_e^{\left( n \right)} = {r_{\min }} - {{n'_n}}\Delta \sin {\theta _{\max }} + \frac{{{{\left( {{{n'_n}}} \right)}^2}{\Delta ^2}{{\cos }^2}{\theta _{\max }}}}{{2{r_{\min }}}} $,
where $ n'_n = \frac{{2(n-1) - N + 1}}{2} $, $\forall n \in \left\{ {1,2, \cdots ,N} \right\}$.
Subsequently, each UE feeds back the subcarrier index with the maximum received power to the BS.
The BS can determine the estimated angle for each UE based on these feedback utilizing formulas \eqref{8} and \eqref{9}.
In the subsequent stage, beam sweeping is carried out independently for each UE.
Specifically, for UE $k$, the BS employs beamforming akin to trajectory 3 in Fig. \ref{T} to transmit $M$ subcarriers with angles of $ {{\hat \theta }_k} $, spanning distances from $r_{\text{min}}$ to $r_{\text{max}}$ in the form of a straight segment.
This approach facilitates $M$ subcarriers to encompass all possible positions of UE $k$, thus facilitating the estimation of its position similar to the angle estimation stage.
Fig. \ref{power} compares the performance of the scheme proposed in \cite{[3]} under ideal noise-free conditions in spatial stationary and non-stationary channels. The UE is located at $(15$ ${\rm m}, 0^\circ)$ with $ M =2048 $, $ f_c = 100 $ GHz, $ N =512 $, and $ B =6 $ GHz.
In spatially stationary channels, the angle and distance estimates are $0.0099^\circ$ and $14.9889$ m, respectively, with corresponding absolute errors of $0.0099^\circ$ and $0.0111$ m.
In contrast, in spatially non-stationary channels, assuming the user's VR includes the first antenna to the $128$-th antenna of the BS, the angle and distance estimates are $0.0871^\circ$ and $20.0899$ m, respectively, with absolute errors of $0.0871^\circ$ and $5.0899$ m.
Notably, the spatial non-stationary characteristics significantly degrade the performance of this scheme.
Furthermore, from the perspective of normalized signal power, during the angle estimation stage, the power variation near the maximum power in Fig. \ref{power} (a) is quite drastic.
Conversely, during the distance estimation stage in Fig. \ref{power} (b), the power variation near the maximum power is relatively gentle.
Specifically, in spatially non-stationary channels, there are $22$ subcarriers in Fig. \ref{power} (a) with power exceeding half of the peak power, while in Fig. \ref{power} (b), there are $688$ subcarriers with power exceeding half of the peak power.
This indicates that in low signal-to-noise ratio (SNR) situations, the random disturbances of noise power have minimal impact on angle estimation performance but have a significant impact on distance estimation performance.
An extreme example is illustrated by assuming that noise causes subcarriers with power exceeding half of the peak power in an ideal noise-free environment to display equal power in a noisy environment. In such a case, the subcarrier index with the maximum power will be offset by $11$ during the angle estimation stage and by $344$ during the distance estimation stage.

\begin{algorithm}[t]
  \caption{The Proposed Controllable Beam Squint Based Beam Training Method. }
  \label{alg:1}
  \begin{algorithmic}[1]

  \Require  Potential angle range $ \left[ {{\theta _{\min }},{\theta _{\max }}} \right] $, potential distance range  $ \left[ {{r _{\min }},{r _{\max }}} \right] $, the bandwidth $B$, the frequency of each subcarrier $ {f_m} $, $ m \in \left\{ {1,2, \cdots ,M} \right\} $, $  n \in \left\{ {1,2, \cdots ,N} \right\} $.
  \Ensure Estimated locations $  \left( {{{\hat r}_k},{{\hat \theta '}_k}} \right)$, $\forall k \in \left\{ {1,2, \cdots ,K} \right\}  $.

  \Statex // \textbf{Stage \uppercase\expandafter{\romannumeral1}:} Estimating the angle of each UE.

    \State Set $ r_s^{\left( n \right)} = {r_{\min }} - n'_n\Delta \sin {\theta _{\min }} + \frac{{{{\left( {n'_n} \right)}^2}{\Delta ^2}{{\cos }^2}{\theta _{\min }}}}{{2{r_{\min }}}} $ and $ r_e^{\left( n \right)} = {r_{\min }} - {{n'_n}}\Delta \sin {\theta _{\max }} + \frac{{{{\left( {{{n'_n}}} \right)}^2}{\Delta ^2}{{\cos }^2}{\theta _{\max }}}}{{2{r_{\min }}}} $ in \eqref{7},


    \State Received signal: $ y_{k,m}^1 = {\bf{h}}_{k,m}^H{\bf{w}}_{k,m}^1s_{k,m}^1 + n_{k,m}^1 $,

    \For{$ k = 1,2,\cdots,K$}

        \State Calculate the signal power $ \left| {y_{k,m}^1} \right| $, 
        \State Feedbacks the index of maximum power $ m_{max}^1 $
        \State Calculate the estimated angle $ {{\hat \theta }_k} $ by \eqref{8} with $ m_{max}^1 $.

    \EndFor

  \Statex // \textbf{Stage \uppercase\expandafter{\romannumeral2}:} Refining the angle of each UE.

    \For{$ k = 1,2,\cdots,K$}

        \State Partition the $M$ subcarriers into $L$ groups and focus $M$ subcarriers on $L$ points: $ {r_{c,l}} = {r_{\min }}$, $\forall l \in \left\{ {1,2, \cdots ,L} \right\} $, $ {\theta _{c,1}} = \min \left\{ {{\theta _{\max }},{{\hat \theta }_k} + 1^\circ } \right\} $, $ {\theta _{c,L}} = \max \left\{ {{\theta _{\min }},{{\hat \theta }_k} - 1^\circ } \right\} $, and $ {\theta _{c,l}} = {\theta _{c,1}} - \left( {l - 1} \right)\left( {{\theta _{c,1}} - {\theta _{c,L}}} \right)/\left( {L - 1} \right) $.
        \State Received signal: $ y_{k,m}^2 = {\bf{h}}_{k,m}^H{\bf{w}}_{k,m}^2s_{k,m}^2 + n_{k,m}^2 $,
        \State Calculate the total signal power for each group: $ y_{k,l}^2 = \sum\nolimits_{m = \left( {l - 1} \right)L/M + 1}^{lL/M} {\left| {y_{k,m}^2} \right|}   $ and the index of the group with the highest signal power $ l_{max} $,
        \State Obtain the refined the angle estimation: $ {{\hat \theta '}_k} = {\theta _{c,l_{\max }}} $.

    \EndFor

  \Statex // \textbf{Stage \uppercase\expandafter{\romannumeral3}:} Estimating the distance of each UE.
    \State ${r_{{\rm{start}}}^0} = {r_{\min }}$, ${r_{{\rm{end}}}^0} = {r_{\max }}$, $j=1$.

    \For{$ k = 1,2,\cdots,K$}
        \While{$ j \leq J $ and $ {r_{{\rm{end}}}^j} - {r_{{\rm{start}}}^j} \geq \varepsilon $}

            \State Partition the $M$ subcarriers into $L_2$ groups and focus $M$ subcarriers on $L_2$ points: $ \theta _{c,l}^j = {{\hat \theta '}_k} $, $ r_{c,l}^j = r_{{\rm{start}}}^j + (l - 1)(r_{{\rm{end}}}^j - r_{{\rm{start}}}^j)/({L_2} - 1) $, $ \forall l \in \left\{ {1,2, \cdots ,{L_2}} \right\} $,
            \State Received signal: $ y_{k,m}^3 = {\bf{h}}_{k,m}^H{\bf{w}}_{k,m}^3s_{k,m}^3 + n_{k,m}^3 $,
            \State Calculate the total signal power for each group: $ y_{k,l}^3 = \sum\nolimits_{m = \left( {l - 1} \right)L_2/M + 1}^{lL_2/M} {\left| {y_{k,m}^3} \right|}   $ and the index of the group with the highest signal power $\hat l_{max}^j $,
            \State $j = j + 1$,
            \State $ r_{{\rm{start}}}^j = r_{c,\hat l_{\max }^{(j-1)} - 1}^{j - 1} $ and $ r_{{\rm{start}}}^j = r_{c,\hat l_{\max }^{(j-1)} + 1}^{j - 1} $,

        \EndWhile
        \State Obtain the distance estimation: $ {{\hat r}_k} = r_{c,\hat l_{\max }^J}^J $
    \EndFor

  \end{algorithmic}
\end{algorithm}

Inspired by \cite{[3]}, we integrate the beam squint-based user localization scheme with beam training strategies to further harness the potential of near-field controllable beam squint in user localization.
Our proposed user localization scheme unfolds in three stages, which is presented in {\textbf{{Algorithm}} \ref{alg:1}}.
Note that the first stage of {\textbf{{Algorithm}} \ref{alg:1}} directly adopts the angle estimation strategy in \cite{[3]}, while the second and third stages are our proposed subcarrier grouping strategies.

In the initial stage, akin to the procedure outlined in \cite{[3]}, we derive preliminary angle estimates for each UE, denoted as $ {{\hat \theta }_k}$, $\forall k \in \left\{ {1,2, \cdots ,K} \right\} $.
Subsequently, in the second stage, we refine the angles of each UE.
Without lost of generality, let us take UE $k$ as an example.
Specifically, we organize $M_s$ subcarriers into groups, with each subcarrier within a group focused on a singular point, while subcarriers from distinct groups target different positions.
For simplicity, we assume that $L = M / M_s$ is an integer.
Following this, the PSs and TTDs are recalibrated to
$ {\phi _{n,l}} = \frac{{{f_{1 + \left( {l - 1} \right){M_s}}}r_{c,l}^{\left( n \right)}}}{c} $,
$ {{\tilde f}_m} = {f_m} - {f_{1 + \left( {l - 1} \right){M_s}}} $, and
$ {t_n} = \frac{{{f_{l{M_s}}}r_{c,l}^{\left( n \right)}}}{{c\left( {{f_{{M_s}}} - {f_1}} \right)}} - \frac{{{\phi _n}}}{{\left( {{f_{{M_s}}} - {f_1}} \right)}} $,
where $  l \in \left\{ {1,2, \cdots ,L} \right\} $ is the index of the group, $ r_{c,l}^{\left( n \right)} = {r_{c,l}} - n\Delta \sin {\theta _{c,l}} + \frac{{{n^2}{\Delta ^2}{{\cos }^2}{\theta _{c,l}}}}{{2{r_{c,l}}}} $, and $ ({r_{c,l}}, {\theta _{c,l}}) $ is the location focused by the $l$-th group.
In this stage, we set $ {r_{c,l}} = {r_{\min }}$, $\forall l \in \left\{ {1,2, \cdots ,L} \right\} $, $ {\theta _{c,1}} = \min \left\{ {{\theta _{\max }},{{\hat \theta }_k} + 1^\circ } \right\} $, $ {\theta _{c,L}} = \max \left\{ {{\theta _{\min }},{{\hat \theta }_k} - 1^\circ } \right\} $, and $ {\theta _{c,l}} = {\theta _{c,1}} - \left( {l - 1} \right)\left( {{\theta _{c,1}} - {\theta _{c,L}}} \right)/\left( {L - 1} \right) $.
On the UE side, UE $k$ adds up the power of subcarrier signals within the same group and feeds back the index of the group with the highest power to the BS.
In this way, the BS can obtain a more precise angle estimation $ {{\hat \theta '}_k} $ based on received feedback.

In the third stage, we perform an iterative fine estimation of the distance for UE $k$.
Similar to the second stage, we partition the $M$ subcarriers into $L_2$ groups and initialize the start and end distances of the initial beam sweeping to $ {r_{{\rm{start}}}^0} = {r_{\min }} $ and $ {r_{{\rm{end}}}^0} = {r_{\max }} $, respectively.
For the first iteration of beam sweeping, we set $ {\theta _{c,l}^1} = {{\hat \theta '}_k} $, and $ {r_{c,l}^1} = {r_{{\rm{start}}}^{0}} + (l - 1)({r_{{\rm{end}}}^1} - {r_{{\rm{start}}}^{0}})/({L_2} - 1) $, $ \forall l \in \left\{ {1,2, \cdots ,{L_2}} \right\} $.
Consequently, the $M$ subcarriers will be focused on $L_2$ points.
In a manner akin to the second stage, UE $k$ will relay feedback on the index $ \hat l_{\max }^1 $ of the group with the highest power to the BS.
For the $j$-th iteration of beam sweeping, the start and end distances of the $j$-th beam sweeping are adjusted to $ r_{{\rm{start}}}^j = r_{c,\hat l_{\max }^{(j-1)} - 1}^{j - 1} $ and $ r_{{\rm{end}}}^j = r_{c,\hat l_{\max }^{(j+1)} + 1}^{j - 1} $, respectively.
All angles of beam sweeping remain set to  $ \theta _{c,l}^j = {{\hat \theta '}_k} $, $ \forall l \in \left\{ {1,2, \cdots ,{L_2}} \right\} $.
Consequently, the beam sweeping encompasses $L_2$ distances $ r_{c,l}^j = r_{{\rm{start}}}^j + (l - 1)(r_{{\rm{end}}}^j - r_{{\rm{start}}}^j)/({L_2} - 1) $.
Following $J$ iterations, we derive the index $ \hat l_{\max }^J $.
Subsequently, the estimated distance of UE $k$ is computed as $ {{\hat r}_k} = r_{c,\hat l_{\max }^J}^J $.
It is worth noting that to ensure the convergence of the third stage, a suboptimal number of iterations under a given threshold $ \varepsilon $ can be simulated in a noise-free environment.
This is because under noise-free conditions, the closer the beamforming points from different subcarriers are to the user, the higher the beamforming gain.
Therefore, convergence can be achieved. 

It is noteworthy that the first stage of our proposed scheme directly incorporates the initial stage of the scheme outlined in \cite{[3]} to attain a preliminary estimate of the angles of UEs.
Subsequently, we amalgamate the concepts of controllable beam squint and beam training to refine angle estimates in the second stage and distance estimates in the third stage, respectively.
Our proposed solution delves deeper into the potential of controllable beam squint, and its performance comparison will be elucidated in Section \ref{Simulation Results}.

\subsection{Proposed DL Methods}

\begin{figure}
  \centering
  \subfigure[{{The ConvNeXt based user localization scheme.}}] {\includegraphics[width=2.6in]{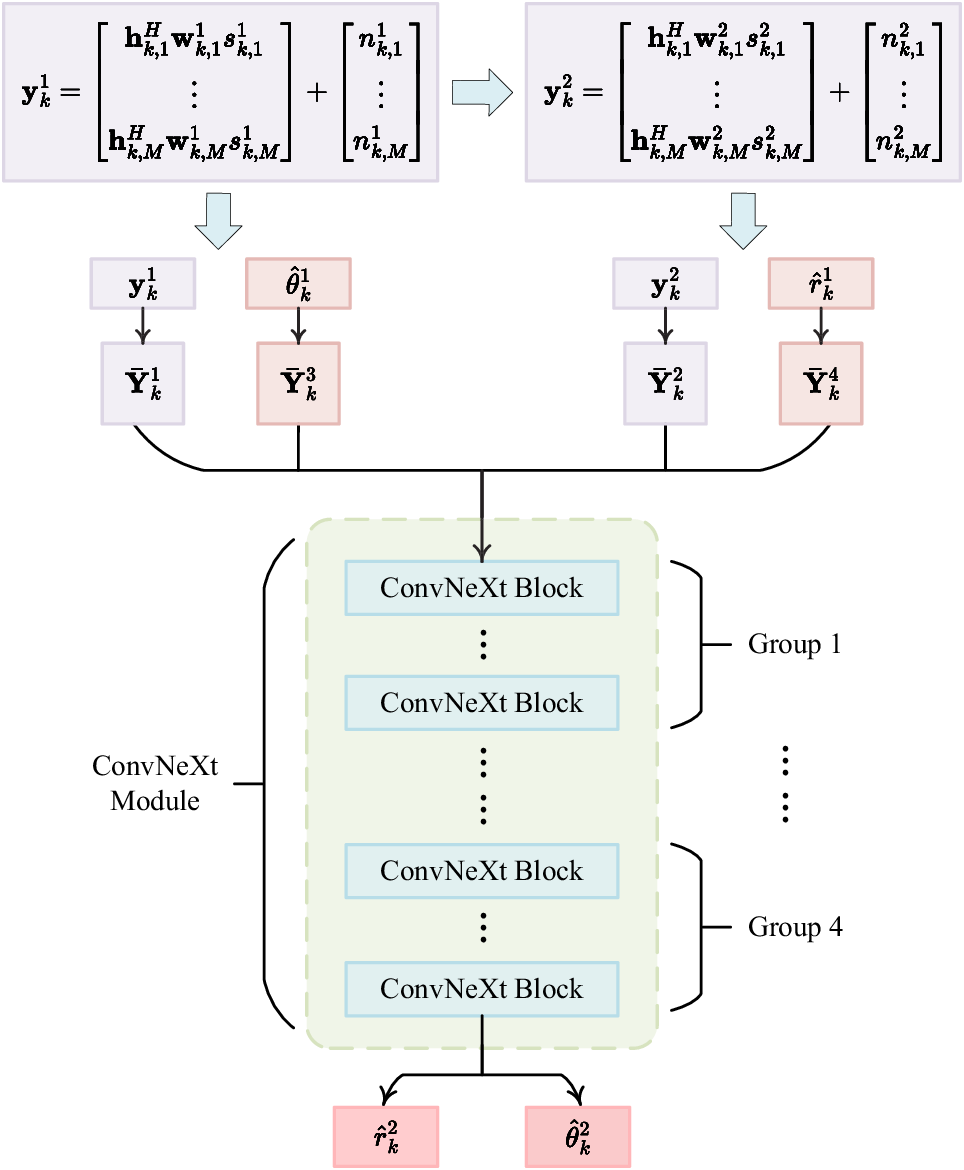}}
  \subfigure[{{The model of the ConvNeXt block.}}] {\includegraphics[width=2.6in]{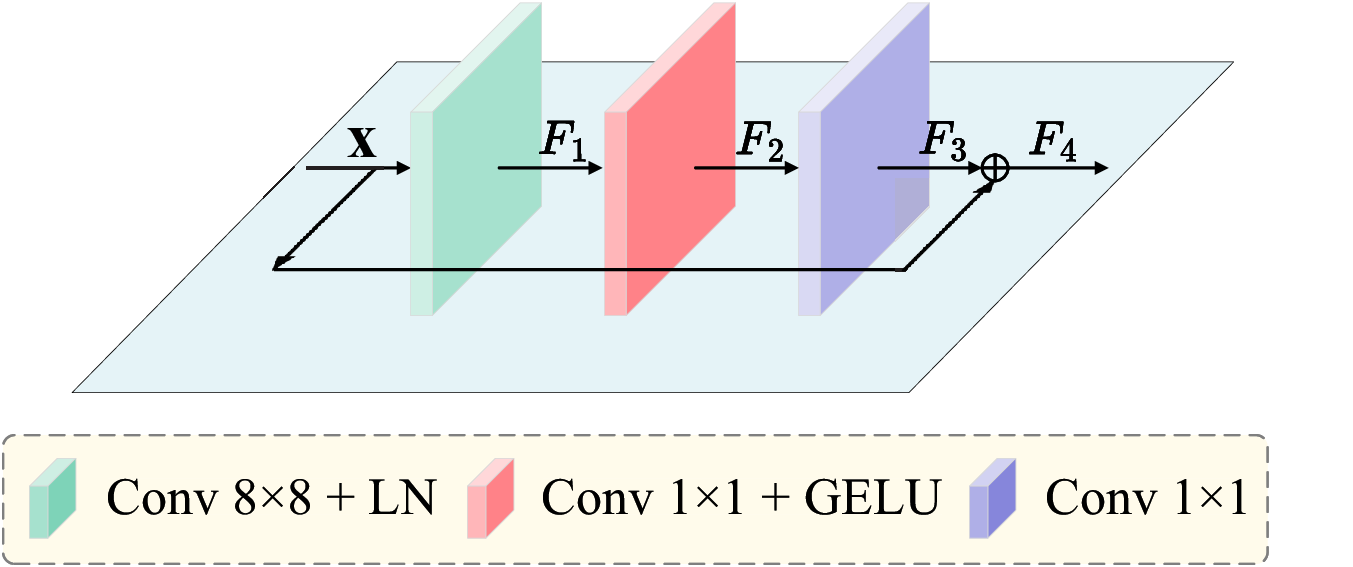}}
  \caption{The proposed ConvNeXt based user localization scheme. It is worth noting that the proposed ConvNeXt based method is suitable for multi-user localization scenarios. Without loss of generality, Fig. 5 only takes user $k$ as an example.}
  \label{DL}
\vspace{-0.6cm}
\end{figure}

In this subsection, we introduce a user localization scheme based on deep learning.
This scheme leverages the output of the user localization scheme in \cite{[3]} as input and utilizes the ConvNeXt network in \cite{[8]} as the primary model to generate high-precision user position estimations.
Specifically, the scheme in \cite{[3]} utilizes the characteristics of near-field spherical waves and beam squint effects to obtain estimates of angle and distance based on the difference in received power among subcarriers.
However, due to the influence of spatial non-stationary characteristics and noise, this scheme fails to effectively exploit the spherical wave characteristics and beam splitting effect, resulting in a significant difference between the expected and actual performance.
Therefore, our proposed approach aims to employ the ConvNeXt network, utilizing its expansive perceptual field of view afforded by large convolutional kernels, to learn noise characteristics and the connections and mappings between the rough estimation information and different subcarriers.
By doing so, it mitigates the impact of spatial non-stationary characteristics and noise on received power.
Consequently, the proposed scheme can delve deeper into the precise user's angle and distance information by leveraging the beam focusing induced by spherical wave characteristics and the controllable beam squint effect arising from large bandwidth. 

As shown in Fig. \ref{DL}, the proposed ConvNeXt based user localization scheme is executed at the BS based on user feedback and primarily comprises the following components.

\subsubsection{Input}
The proposed ConvNeXt method takes the input and output of the beam squint based user localization scheme in \cite{[3]} as its inputs.
Specifically, after the execution of the scheme in \cite{[3]}, the BS can preliminarily derive the approximate angle and distance estimation of UE $k$, denoted by $ \hat \theta _k^1 $ and $\hat r_k^1 $, respectively.
Meanwhile, UE $k$ feeds back the received signals $ {\bf{y}}_k^1 \in {\mathbb{C}^{{M} \times 1 }} $ in the angle estimation stage and $ {\bf{y}}_k^2 \in {\mathbb{C}^{{M} \times 1 }} $ in the distance estimation stage to the BS.
Next, we appropriately preprocess the data to prepare it for input into the DL module.
Specifically, the signals $ {\bf{y}}_k^1 \in {\mathbb{C}^{{M} \times 1 }} $ and the signal $ {\bf{y}}_k^2 \in {\mathbb{C}^{{M} \times 1 }} $ are first reorganized into matrices $ {\bf{Y}}_k^1 \in {\mathbb{R}^{{M} \times 2 }} $ and $ {\bf{Y}}_k^2 \in {\mathbb{R}^{{M} \times 2 }} $, respectively.
Subsequently, these matrices are further restructured into matrices $ {\bf{\bar Y}}_k^1 \in {^{{M_1} \times {M_2}}} $ and $ {\bf{\bar Y}}_k^2 \in {^{{M_1} \times {M_2}}} $, respectively.
Note that we ensure that $M_1>M_2$, $M_1M_2=M$, and the difference between $M_1$ and $M_2$ is as small as possible to ensure that $ {\bf{\bar Y}}_k^1 $ and $ {\bf{\bar Y}}_k^2  $ can be considered as two two-dimensional images.
In addition, we reconstruct matrices $ {\bf{\bar Y}}_k^3 \in {^{{M_1} \times {M_2}}}  $ and $ {\bf{\bar Y}}_k^4 \in {^{{M_1} \times {M_2}}}  $ from $ \hat \theta _k^1 $ and $\hat r_k^1 $, respectively.
Then, the matrices $ {\bf{\bar Y}}_k^1 $, $ {\bf{\bar Y}}_k^2  $, $ {\bf{\bar Y}}_k^3 $, and $ {\bf{\bar Y}}_k^4  $ serve as the input of the proposed ConvNeXt method.

\subsubsection{ConvNeXt Block Structure}

We denote \emph{Convolution}, \emph{Layer Normalization}, and \emph{Gaussian Error Linear Unit} by Conv, LN, and GELU, respectively.
As shown in Fig. \ref{DL} (b), the ConvNeXt block comprises three components.
The first component consists of a Conv layer with a kernel size of $8 \times 8$ (represented as Conv $8 \times 8$) and a LN layer.
The second component encompasses a Conv layer with a kernel size of $1 \times 1$ (represented as Conv $1 \times 1$) and a GELU layer.
The third component is a Conv layer with a kernel size of $1 \times 1$.
Moreover, the recursive functions of the three components are denoted as $ f_1 $, $ f_2 $, and $ f_3 $, respectively.
Then, the recursive relation of the ConvNeXt block can be expressed as
$ {{ F}_1} = {f_1}\left( {\bf{x}} \right) $, $ {{ F}_2} = {f_2}\left( {{{ F}_1}} \right) $, $ {{ F}_3} = {f_3}\left( {{{ F}_2}} \right) $, and $ {{ F}_4} = {{\cal F}}\left( {\bf{x}} \right)  = {{ F}_3} \oplus {\bf{x}} $,
where $ {\bf{x}} $ represents the input of the ConvNeXt block, $ \oplus $ denotes the feature fusion, and $ {{\cal F}}  $ denotes the recursive functions of the ConvNeXt block.
In addition, the number of channels in the Conv layer of the three components is configured as $ C $, $3C $, and $C$, respectively.

\subsubsection{ConvNeXt Structure}

As shown in Fig. \ref{DL}, the proposed ConvNeXt-based user localization scheme utilizes multiple ConvNeXt blocks.
Specifically, we organize all ConvNeXt blocks into four groups, each comprising $E_1$, $E_2$, $E_3$, and $E_4$ ConvNeXt blocks.
Additionally, we represent the number of channels in the first component of the ConvNeXt blocks within each of the four groups as $ C_1$, $ C_2$, $ C_3$, and $ C_4$, respectively.
Then, the recursive relation of the $ i $-th ConvNeXt group is as follows: ($\forall i \in \left\{ {1,2,3,4} \right\} $)
$ {\cal X}_i\left( {\bf{x}} \right) = {{\cal F}_{{E_i}}}\left( { \cdots \left( {{{\cal F}_3}\left( {{{\cal F}_2}\left( {{{\cal F}_1}\left( {\bf{x}} \right)} \right)} \right)} \right)} \right) \buildrel \Delta \over = {{\cal F}_{{E_i}}} \circ  \cdots  \circ {{\cal F}_2} \circ {{\cal F}_1}\left( {\bf{x}} \right), $
where $ {{\cal F}_{{j}}}, j \in \left\{ {1,2,\cdots,E_i} \right\} $, $\forall j \in \left\{ {1,2,\cdots,E_i} \right\}$, denotes the $ j $-th ConvNeXt block in the $ i $-th ConvNeXt group, $ \circ $ denotes a function composition, $ {\bf{x}} $ is the input of the $ i $-th ConvNeXt group, and $ {\cal X}_i\left( {\bf{x}} \right) $ is the output of the $ i $-th ConvNeXt group.
Ultimately, the recursive relation of the ConvNeXt module is
$ {\cal T}\left( {\bf{x}} \right) = {{\cal X}_4}\left( {{{\cal X}_3}\left( {{{\cal X}_2}\left( {{{\cal X}_1}\left( {\bf{x}} \right)} \right)} \right)} \right) \buildrel \Delta \over = {{\cal X}_4} \circ {{\cal X}_3} \circ {{\cal X}_2} \circ {{\cal X}_1}\left( {\bf{x}} \right), $
where $ {\cal T}\left( {\bf{x}} \right): $ $ {\mathbb{R}^{4 \times {M_1} \times {M_2}}} \to {\mathbb{R}^{2 \times 1}} $ is the mapping function for the ConvNeXt module.
In other words, the ConvNeXt module constructs a mapping from the data $ \left\{ {{\bf{\bar Y}}_k^1,{\bf{\bar Y}}_k^2,{\bf{\bar Y}}_k^3,{\bf{\bar Y}}_k^4} \right\} $ to the user`s location estimation $ \left\{ {\hat r_k^2,\hat \theta _k^2} \right\} $.
It is worth noting that the proposed ConvNeXt based method is suitable for multi-user localization scenarios.
For the sake of clarity, we exemplify the process with user $k$, without loss of generality.

\subsubsection{Loss Function}

The purpose of our proposed ConvNeXt based method is to maximize user localization performance.
Therefore, based on the commonly used root mean square error (RMSE) metric in sensing and localization, the loss function can be expressed as
$ {\cal L} =  \sqrt {{{\left( {\hat r_k^2 - {r_k}} \right)}^2}}  + \sqrt {{{\left( {\hat \theta _k^2 - {\theta _k}} \right)}^2}} $.


\subsubsection{Computational Complexity}

The computational complexity of the running phase in the ConvNeXt is given by $ \mathcal{O}({C_1} {E_1^2} K^2 F_1^2) + \mathcal{O}({C_2} {E_2^2} K^2 F_2^2) + \mathcal{O}({C_3} {E_3^2} K^2 F_3^2) + \mathcal{O}({C_4} {E_4^2} K^2 F_4^2)$,
where $F_i$ is the input dimension of the $i$-th ConvNeXt group.
To reduce complexity, downsampling is performed before each ConvNeXt group. Specifically, before the first ConvNeXt group, there is a Conv layer with a kernel size of $4 \times 4$, along with a LN layer, is applied.
Before the second, third, and fourth ConvNeXt groups, a Conv layer with a kernel size of $2 \times 2$ and a stride size of $2 \times 2$, along with a LN layer, is employed.
Consequently, the input dimensions are $ F_1 = 16 $, $ F_2 = 8 $, $ F_3 = 4 $, and $ F_4 = 2 $.

\section{Simulation Results}\label{Simulation Results}

In this section, the performance of the proposed algorithms are evaluated through numerical simulations.
We begin by examining the CRB of user localization in a wideband XL-MIMO system.
Our focus lies in investigating the impact of the number of subcarriers and bandwidth on CRB.
This analysis aims to offer insights for subsequent system settings.
Subsequently, we evaluate the potential of the methods based on controllable beam squint.
Specifically, we analyze and compare the performance of controllable beam squint based scheme in \cite{[3]} with our proposed beam training approach, leveraging controllable beam squint assistance, under various conditions.
Finally, we evaluate the user localization performance of the ConvNeXt based solution.

\vspace{-0.3cm}
\subsection{Simulation Setup}

Unless otherwise specified, the following setup is employed throughout the simulations.
We consider a wideband XL-MIMO system, where the number of antennas at the BS is $ N =512 $, the central carrier frequency is $ f_c  = 100$ GHz, the number of the subcarriers is $ M = 2048 $, and the bandwidth is $ B = 6 $ GHz.
We assume that the UEs are distributed around the BS such that $ r \sim {\mathcal{U}}[ 5$ ${\rm{m}},50$ ${\rm{m}} ] $ and $ \theta  \sim {\mathcal{U}}\left[ { - {\pi  \mathord{\left/  {\vphantom {\pi  3}} \right.  \kern-\nulldelimiterspace} 3},{\pi  \mathord{\left/  {\vphantom {\pi  3}} \right.  \kern-\nulldelimiterspace} 3}} \right] $.
We utilize the root mean square error (RMSE) to quantify the user localization performance.
The RMSE of the angle, distance, and location for UE $k$ can be defined as ${\rm{RMS}}{{\rm{E}}_{k,\theta }} = \sqrt {{{\sum\nolimits_{i = 1}^{{N_{{\rm{iter}}}}} {\left( {{{\hat \theta }_{k,i}} - {\theta _{k,i}}} \right)} } \mathord{\left/
 {\vphantom {{\sum\nolimits_{i = 1}^{{N_{{\rm{iter}}}}} {\left( {{{\hat \theta }_{k,i}} - {\theta _{k,i}}} \right)} } {{N_{{\rm{iter}}}}}}} \right.
 \kern-\nulldelimiterspace} {{N_{{\rm{iter}}}}}}}  $, $ {\rm{RMS}}{{\rm{E}}_{k,r}} = \sqrt {{{\sum\nolimits_{i = 1}^{{N_{{\rm{iter}}}}} {\left( {{{\hat r}_{k,i}} - {r_{k,i}}} \right)} } \mathord{\left/
 {\vphantom {{\sum\nolimits_{i = 1}^{{N_{{\rm{iter}}}}} {\left( {{{\hat r}_{k,i}} - {r_{k,i}}} \right)} } {{N_{{\rm{iter}}}}}}} \right.
 \kern-\nulldelimiterspace} {{N_{{\rm{iter}}}}}}}  $, and ${\rm{RMS}}{{\rm{E}}_{k,{\rm{2D}}}} = \sqrt {{{\left[ {\sum\nolimits_{i = 1}^{{N_{{\rm{iter}}}}} {\left( {{{\hat x}_{k,i}} - {x_{k,i}}} \right) + \sum\nolimits_{i = 1}^{{N_{{\rm{iter}}}}} {\left( {{{\hat y}_{k,i}} - {y_{k,i}}} \right) + } } } \right]} \mathord{\left/
 {\vphantom {{\left[ {\sum\nolimits_{i = 1}^{{N_{{\rm{iter}}}}} {\left( {{{\hat x}_{k,i}} - {x_{k,i}}} \right) + \sum\nolimits_{i = 1}^{{N_{{\rm{iter}}}}} {\left( {{{\hat y}_{k,i}} - {y_{k,i}}} \right) + } } } \right]} {{N_{{\rm{iter}}}}}}} \right.
 \kern-\nulldelimiterspace} {{N_{{\rm{iter}}}}}}} $, respectively, where $\left( {{{\hat x}_{k,i}},{{\hat y}_{k,i}}} \right)$ and $ \left( {{x_{k,i}},{y_{k,i}}} \right) $ are the Cartesian coordinates corresponding to polar coordinates $\left( {{{\hat r}_{k,i}},{{\hat \theta }_{k,i}}} \right)$ and $ \left( {{r_{k,i}},{\theta _{k,i}}} \right) $.
We set $ {{N_{{\rm{iter}}}}} = 5000 $, ensuring that $5000$ independent channel and noise realizations are utilized to conduct Monte Carlo simulations.
Moreover, the signal-to-noise ratio (SNR) is defined as
$ {\rm{SN}}{{\rm{R}}_{k,m}} = {{\mathbb{E}\left\{ {{{\left| {{\bf{h}}_{k,m}^H{{\bf{w}}_{k,m}}} \right|}^2}} \right\}} \mathord{\left/
 {\vphantom {{\mathbb{E}\left\{ {{{\left| {{\bf{h}}_{k,m}^H{{\bf{w}}_{k,m}}} \right|}^2}} \right\}} {\mathbb{E}\left\{ {{{\left| {{n_{k,m}}} \right|}^2}} \right\}}}} \right.
 \kern-\nulldelimiterspace} {\mathbb{E}\left\{ {{{\left| {{n_{k,m}}} \right|}^2}} \right\}}} $.
For brevity, we set $ {\rm{SNR }} =  {\rm{SNR }}_{k,m} $, $ \forall m \in \left\{ {1,2, \cdots ,M} \right\},\forall k \in \left\{ {1,2, \cdots ,K} \right\} $.

We set $ J = 5 $ and $ \varepsilon = 0.5 $ as the number of the iteration and the pre-specified threshold for stopping criterion in {\textbf{{Algorithm}} \ref{alg:1}}, respectively.
In addition, the number of sub-arrays $ N_s $ is set to $4$.
As for the DL module, the learning rate for the ConvNeXt is set to $0.001$.
The number of channels the ConvNeXt block and the number of blocks within each of the four groups will be specified later.
Note that although our proposed schemes are tailored for multi-user scenarios, the location estimation for each UE is independent.
Therefore, without lost of generality, we consider an arbitrary UE in the simulations.

\vspace{-0.3cm}
\subsection{Cram\'er-Rao Bound Performance}


\begin{figure}
\centering
\subfigure[The root CRB of angle.] {\includegraphics[width=2.5in]{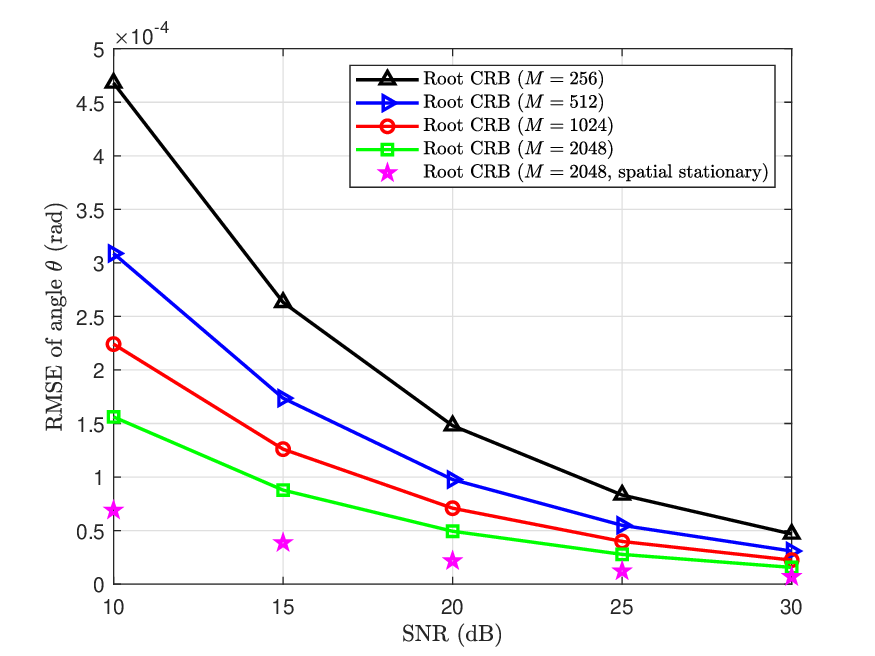}}
\subfigure[The root CRB of distance.] {\includegraphics[width=2.5in]{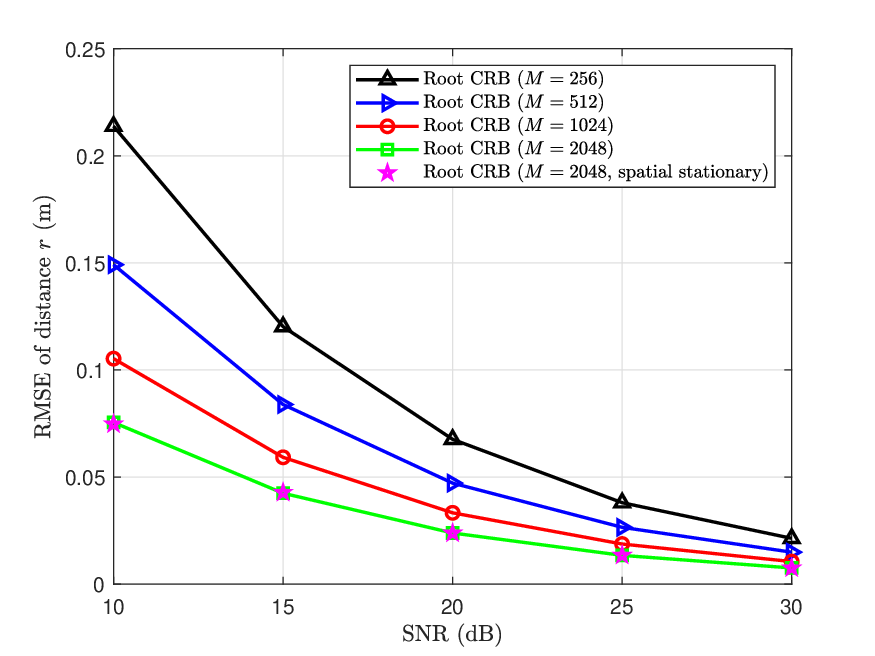}}
\caption{The root CRB of angle and distance for different number of subcarriers as a function of SNR.
}
\label{RCRB_M}
\vspace{-0.6cm}
\end{figure}

Fig. \ref{RCRB_M} (a) shows the root CRB of angle versus the SNRs for different number of subcarriers.
The first observation is that as the SNR increases, the root CRB of angle decreases across different subcarrier numbers. 
Moreover, as the SNR increases, the rate of decrease in the root CRB of angle at different subcarrier numbers gradually slows down.
This is because the higher SNR leads to a gradual reduction in noise power, thereby diminishing its impact on angle estimation.
Secondly, it is observed that as the number of subcarriers increases, the root CRB of the angle decreases.
For instance, at a SNR of $10$ dB, the root CRB of angle with $2048$ subcarriers is $85.26\%$ lower than that with $256$ subcarriers.
This reduction is attributed to the increase in the number of angles covered by beamforming as the number of subcarriers increases.
Consequently, more angles are explored within the same angle range during a single beam sweeping, resulting in a lower root CRB.
It is worth noting that the improvement in localization performance due to an increase in the number of subcarriers is more pronounced in low SNR scenarios than in high SNR scenarios.
For instance, at a SNR of $30$ dB, the root CRB of angle with $2048$ subcarriers decreases by $85.21\%$ compared to that with $256$ subcarriers, which is slightly less than the reduction observed at a SNR of $10$ dB.
Furthermore, we observe that when the number of subcarriers is $2048$, the root CRB of angle in spatially stationary channels is notably lower than that in spatially non-stationary channels.
This discrepancy arises from the spatial non-stationary characteristics, which can diminish the spatial resolution of the BS.

Fig. \ref{RCRB_M} (b) depicts the root CRB of distance versus the SNRs for different number of subcarriers.
The trend regarding the influence of the SNR and the number of subcarriers on the root CRB of distance aligns with their impact on the root CRB of the angle in Fig. \ref{RCRB_M} (a).
Notably, the impact of spatially non-stationary characteristics on the root CRB of distance is minimal.
When the number of subcarriers is $2048$, the difference between the root CRB of the distance in spatially stationary channels and spatially non-stationary channels is negligible, with both being approximately equal.
The primary reason for this observation lies in the nature of the angle and distance information contained in the received signal \eqref{10}.
The angle information originates from the phase in the channel, specifically the array response vector in \eqref{3}.
Under spatially non-stationary characteristics, only a subset of antennas serves a specific user, rendering only a portion of the array response vector valuable.
Conversely, the distance information is primarily derived from the channel fading component, represented by $ \beta_{k,m} $ in \eqref{4}.
The influence of spatially non-stationary characteristics on this component is much smaller compared to its impact on the angle information contained in the array response vector.
This disparity is also reflected in the calculation of partial derivatives of the ideal signal with respect to angle and distance, as shown in \eqref{a1}. 


\begin{figure}
\centering
\subfigure[The root CRB of angle.]{\includegraphics[width=2.6in]{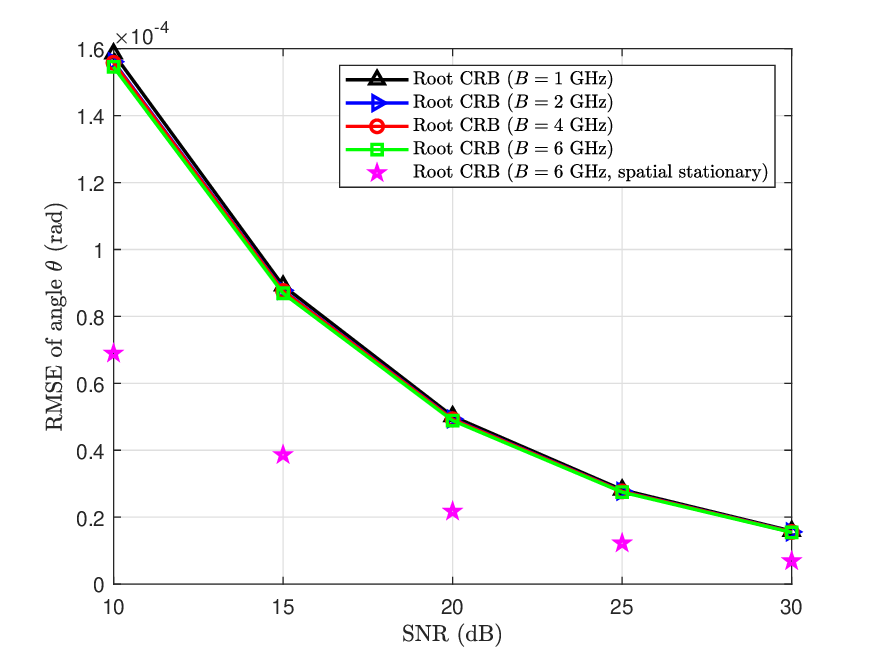}}
\subfigure[The root CRB of distance.]{\includegraphics[width=2.6in]{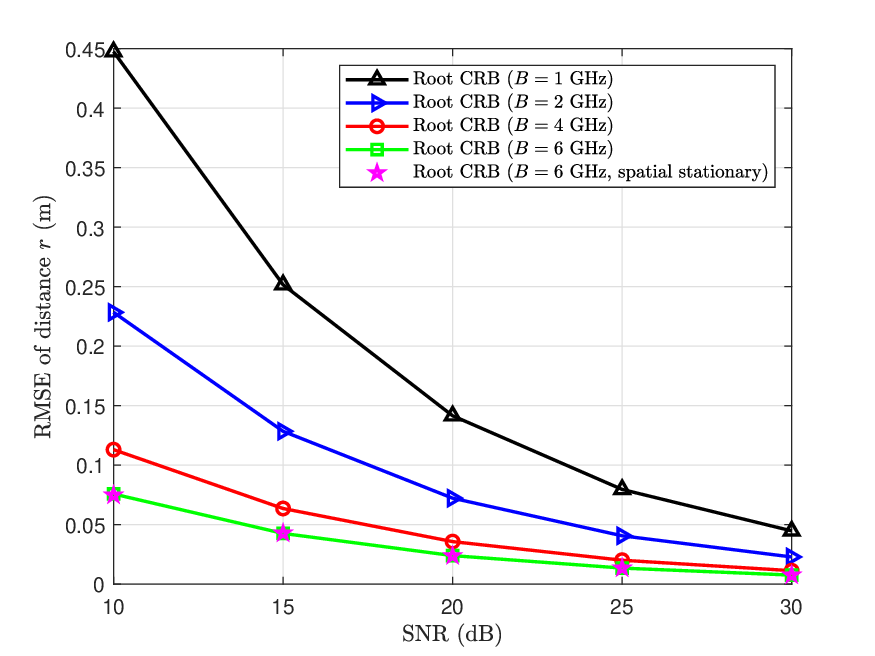}}
\caption{The root CRB of angle and distance for different bandwidths against SNR.
}
\label{RCRB_B}
\vspace{-0.8cm}
\end{figure}

Fig. \ref{RCRB_B} (a) illustrates the root CRB of angle as a function of the SNRs for different bandwidths, with the number of subcarriers set to $2048$.
As the SNR increases, the root CRB of angle at different bandwidths decreases, similar to the trend observed in Fig. \ref{RCRB_M} (a).
Additionally, the influence of spatially non-stationary characteristics on the root CRB of angles mirrors the pattern shown in Fig. \ref{RCRB_M} (a).
Moreover, it is worth noting that increasing bandwidth has negligible impact on the root CRB of the angle.
The rationale lies in the significant difference in power among different subcarrier signals near the maximum subcarrier power at the UE side, as depicted in Fig. \ref{power} (a).
This implies that even with a small subcarrier spacing, the power of subcarriers directed towards the user's angle remains much higher than that of adjacent subcarriers directed towards other angles.
Consequently, the effect of bandwidth on the root CRB of the angle is minimal.
Moreover, Fig. \ref{RCRB_B} (b) displays the root CRB of distance as a function of SNRs for different bandwidths.
The influence of spatially non-stationary characteristics and SNR on the root CRB of distance aligns with Fig. \ref{RCRB_M} (b), reaffirming the conclusion drawn from Fig. \ref{RCRB_M} (b).
However, unlike Fig. \ref{RCRB_B} (a), increasing bandwidth results in a decrease in the root CRB of distance.
This is attributed to the relatively small power difference between different subcarrier signals near the maximum subcarrier power at the user side, as illustrated in Fig. \ref{power} (b).
In essence, the power of beamforming directed towards subcarriers near the user's distance is only slightly higher than that directed towards subcarriers at adjacent distances.
With the same number of subcarriers, widening the bandwidth increases the subcarrier spacing, leading to a larger power difference between subcarriers directed towards the user's location and those towards adjacent locations.
Consequently, this further reduces the root CRB of the distance, resulting in improved theoretical distance estimation performance.

\vspace{-0.3cm}
\subsection{User Localization}


\begin{figure}
\centering
\subfigure[The RMSE comparison for angle estimation.] {\includegraphics[width=2.6in]{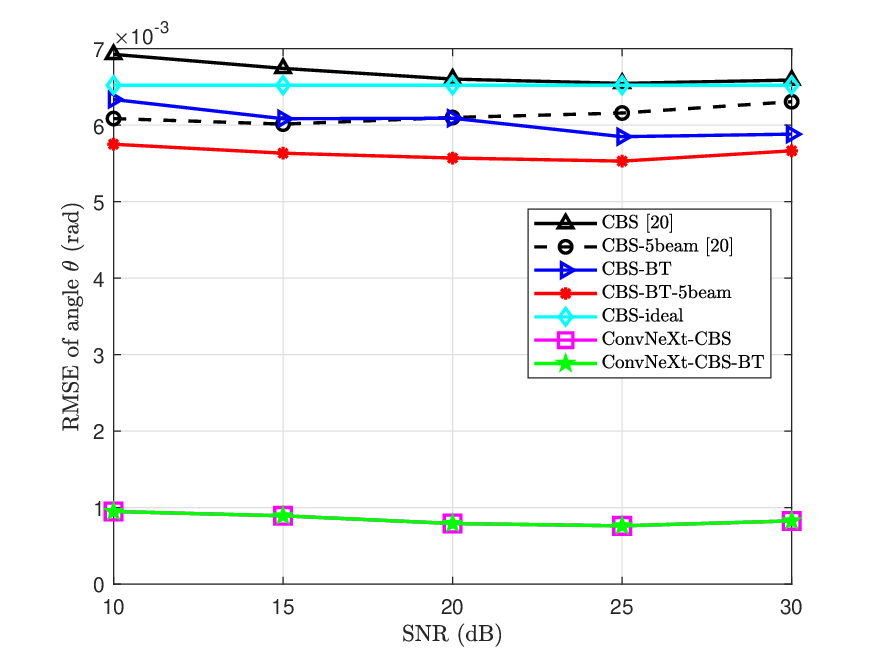}}
\subfigure[The RMSE comparison for distance estimation.] {\includegraphics[width=2.6in]{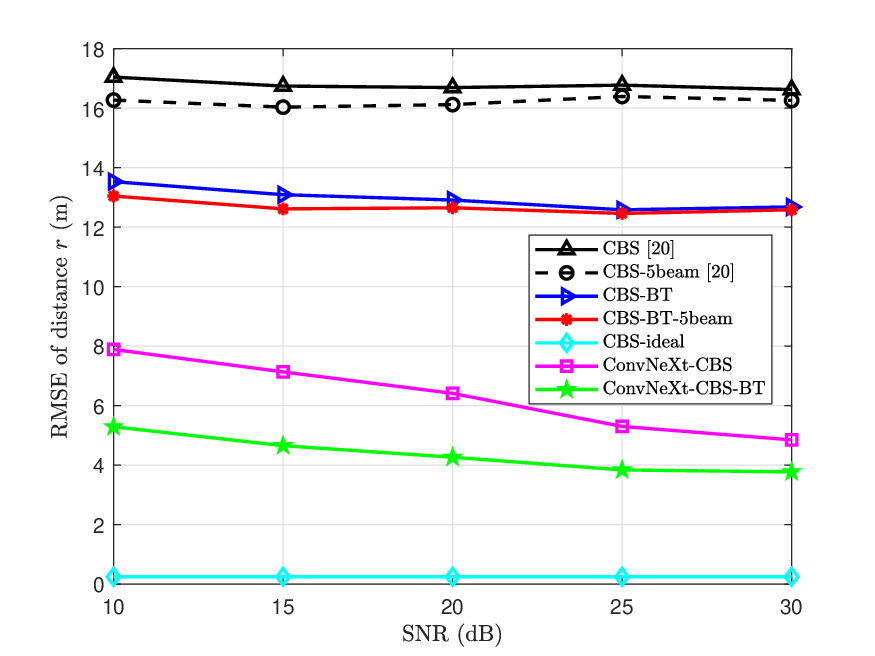}}
\caption{{{The RMSE performance comparison of the CBS method and CBS-BT method under different conditions. }}
}
\label{CBS_BT}
\vspace{-0.8cm}
\end{figure}

In Fig. \ref{CBS_BT} (a), we compare the RMSE performance for angle estimation between the controllable beam squint-based method in \cite{[3]} and our proposed beam training approach.
For brevity, we abbreviate the controllable beam squint-based scheme in \cite{[3]} as {\emph {CBS}} ($1$ beam sweeping for
angle estimation and $1$ beam sweeping for distance estimation).
The proposed controllable beam squint-based beam training method in {\textbf{{Algorithm}} \ref{alg:1}} is denoted as {\emph {CBS-BT}} ($1$ beam sweeping for the first stage and $1$ beam sweeping for the second stage in {\textbf{{Algorithm}} \ref{alg:1}}).
Firstly, it is evident that CBS-BT outperforms CBS in angle estimation.
Specifically, at a SNR of $20$ dB, the angle RMSE of CBS-BT is reduced by $7.75\%$ compared to CBS.
Moreover, we observe that the performance of CBS and CBS-ideal (in an ideal noise-free environment) tends to converge when the SNR exceeds $20$ dB.
This convergence is due to the significant power difference among different subcarriers near the maximum subcarrier power during the angle estimation stage, as illustrated in Fig. \ref{power} (a).
Therefore, under high SNR scenarios, the influence of noise on the angle estimation error of CBS diminishes.
Furthermore, despite both CBS and CBS-BT having the same beam sweeping overhead during the angle estimation stage, i.e. CBS-$5$beam ($4$ beams sweeping for angle estimation and $1$ beam sweeping for distance estimation) and CBS-BT-$5$beam ($2$ beams sweeping for the first stage and $2$ beams sweeping for the second stage in {\textbf{{Algorithm}} \ref{alg:1}}), the proposed CBS-BT still exhibits superior angle estimation performance.
Specifically, at a SNR of $20$ dB, the angle RMSE of CBS-BT-$5$beam is reduced by $8.68\%$ compared to CBS-$5$beam.
Moreover, noteworthy is the superiority of CBS-BT over CBS-ideal, underscoring the feasibility and advantages of leveraging controllable beam squint to group subcarriers and refine angle estimation after initial angle approximation.

In Fig. \ref{CBS_BT} (b), we compare the RMSE performance for distance estimation between the CBS and our proposed CBS-BT under different conditions.
Primarily, it is evident that CBS-BT outperforms CBS in distance estimation.
For instance, at a SNR of $10$ dB, the distance RMSE of CBS-BT is $13.53$ m, compared to $17.05$ m for CBS, indicating a $20.62\%$ reduction in distance RMSE with CBS-BT compared to CBS.
Additionally, the distance estimation performance of CBS-BT-$5$beam surpasses that of CBS-$5$beam.
Specifically, at a SNR of 10 dB, the distance RMSE of CBS-BT-$5$beam is $13.53$ m, whereas for CBS-$5$beam, it is $16.27$ m, representing a $19.82\%$ decrease in distance RMSE with CBS-BT-$5$beam compared to CBS-$5$beam.
It is noteworthy that at this point, both approaches exhibit notable distance estimation errors.
This discrepancy primarily arises from the heavy reliance of both schemes on accurate angle estimation, with noise having a relatively minor impact on their performance.
Upon replacing the angle estimation methods of both schemes with the ConvNeXt scheme proposed in this paper, significant reductions in distance estimation errors were observed.
Specifically, at a SNR of $30$ dB, ConvNeXt-CBS ($4$ beams sweeping for angle estimation based on ConvNeXt and $1$ beam sweeping for distance estimation based on CBS) exhibits a distance RMSE of $4.85$ m, representing a $70.20\%$ decrease compared to CBS-$5$beam.
Similarly, the distance RMSE for ConvNeXt-CBS-BT ($4$ beams sweeping for angle estimation based on ConvNeXt and up to $5$ beams sweeping for distance estimation based on CBS-BT) is $3.77$ m, marking a $71.08\%$ reduction compared to CBS-$5$beam.
Utilizing ConvNeXt for angle estimation, although the distance error of the proposed CBS-BT estimation remains relatively high, it nevertheless provides preliminary information for signal processing algorithms like beam tracking and beamforming.

%

\begin{figure}
\centering
\subfigure[The RMSE performance comparison for angle estimation.] {\includegraphics[width=2.6in]{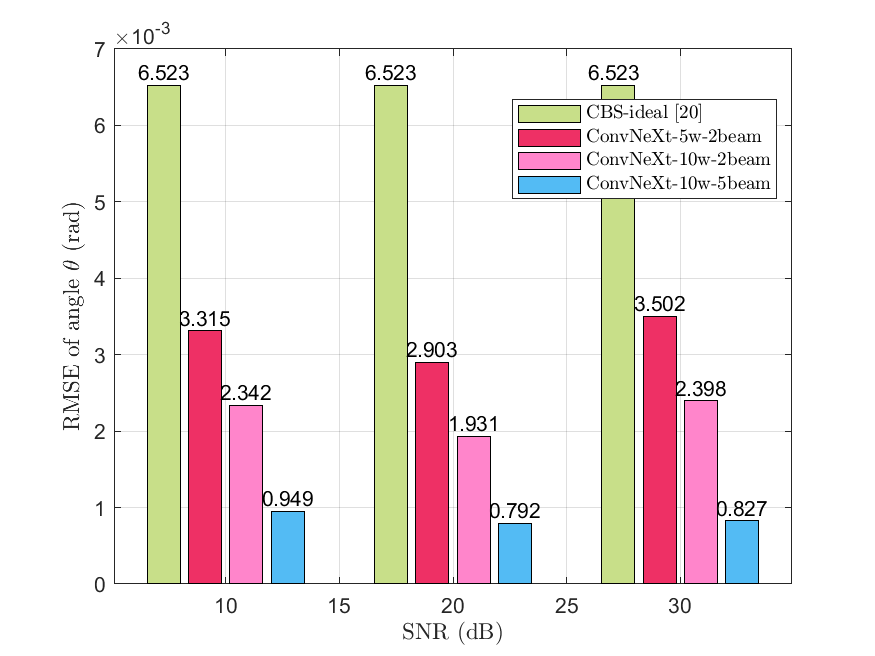}}
\subfigure[The RMSE performance comparison for distance estimation.] {\includegraphics[width=2.6in]{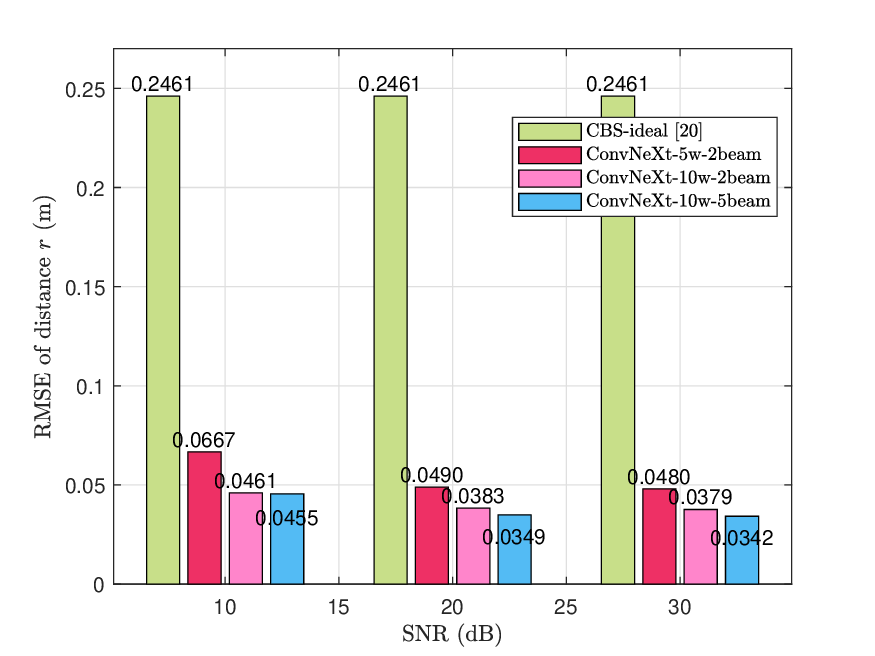}}
\caption{{{The RMSE performance comparison of the CBS method and ConvNeXt method.}}
}
\label{DL_RMSE}
\vspace{-0.8cm}
\end{figure}

Fig. \ref{DL_RMSE} (a) shows a comparison of RMSE performance for angle estimation between the CBS method in an ideal noise-free environment with the proposed ConvNeXt method.
Three different ConvNeXt settings were considered:
(a) ConvNeXt-$5$w-$2$beam: $50000$ training samples, $C_1=C_2=C_4=3$, $C_3=27$, $E_1=128$, $E_2=256$, $E_3=512$, $E_4=1024$,
(b) ConvNeXt-$10$w-$2$beam: $100000$ training samples, $C_1=C_2=C_4=3$, $C_3=27$, $E_1=256$, $E_2=512$, $E_3=1024$, $E_4=2048$,
and (c) ConvNeXt-$10$w-$5$beam:  $100000$ training samples, $C_1=C_2=C_4=3$, $C_3=27$, $E_1=256$, $E_2=512$, $E_3=1024$, $E_4=2048$.
The initial observation is that all three proposed ConvNeXt-based schemes outperform CBS-ideal.
For instance, at a SNR of $10$ dB, the angle RMSE of ConvNeXt-$5$w-$2$beam is $3.31 \times 10^{-3}$ rad, representing a $ 49.18\% $ decrease over CBS-ideal.
Additionally, we notice that enhancing the number of channels and training samples for ConvNeXt could enhance its angle estimation performance.
Specifically, at the same SNR, the angle RMSE of ConvNeXt-$10$w-$2$beam is $2.34 \times 10^{-3}$ rad, marking a $29.35\%$ reduction compared to ConvNeXt-$5 $w-$2 $beam.
Furthermore, with an increased beam sweeping overhead for ConvNeXt, its angle estimation performance further improves.
For instance, at a SNR of $10$ dB, the angle RMSE of ConvNeXt-$10$w-$4$beam is $9.50 \times 10^{-4}$ rad, which is $59.46\%$ lower than ConvNeXt-$10 $w-$2$beam.
This underscores that for ConvNeXt, the angle estimation gains achieved by augmenting the beam sweeping overhead significantly outweigh those from enhancing the model's complexity.
In addition, as the SNR increases, the angle RMSE of ConvNeXt-$5$w-$2$beam and ConvNeXt-$10$w-$2$beam exhibits oscillations.
This phenomenon arises from the predominant influence of the number of random training samples under each SNR on performance, rather than the level of noise.

Fig. \ref{DL_RMSE} (b) depicts the RMSE performance comparison for distance estimation between the CBS-ideal method with the proposed ConvNeXt method.
We can observe that all three proposed ConvNeXt-based solutions outperform CBS-ideal.
In this context, CBS-ideal assumes perfect knowledge of the user's angle during distance estimation and operates without noise.
Notably, the distance estimation accuracy achieved by all three proposed ConvNeXt-based schemes reaches the centimeter level.
For instance, at a SNR of $20$ dB, the distance RMSE of ConvNeXt-$5$w-$2$beam is $0.049$ m, marking an $80.11\%$ reduction compared to the $0.246$ m of CBS-ideal.
Moreover, the RMSE of ConvNeXt-$10$w-$2$beam is $0.038$ m, representing a $21.68\%$ decrease over ConvNeXt-$5$w-$2$beam.
The marginal difference between ConvNeXt-$10$w-$2$beam and ConvNeXt-$10$w-$4$beam can be attributed to their distinct beam sweeping overheads.
While ConvNeXt-$10$w-$4$beam incurs additional costs for angle estimation due to four beams sweeping steps, both models employ a single beam sweeping overhead for distance estimation.
Despite these differences, all three ConvNeXt-based solutions demonstrate remarkable precision in user localization.

\begin{table*}[t]
  \begin{center}
  \renewcommand{\arraystretch}{1.3}

    \caption{{{Comparison of the proposed ConvNeXt with CHISEL.}}}
\label{table}
 \begin{threeparttable}
    \begin{tabular}  {
      c
      c
      c
      c
      c
      c
      c
      c    } 

      \hline
      \hline
         \bf \multirow{2}{*}{{Scenarios}}
        &  \multirow{2}{*}{{ \makecell{ \bf Stationary or\\\bf non-stationary} }}
        & \multirow{2}{*}{\makecell{ \bf Schemes  }}
        & \multirow{2}{*}{\makecell{ \bf FLOPs   }}
        & \multicolumn{4}{c}{{ \bf RMSE of location (m)   }} \\
      \cline{5-8}

      &
      &
      &
      & 0 dB
      & 10 dB
      & 20 dB
      & 30 dB \\
      \hline
      \hline

      \multirow{5}{*}{LoS}
      &  \multirow{5}{*}{Stationary}
      &  CHISEL-1
      & 0.48 M
      & 0.3144
      &  0.2091
      &  0.1864
      &  0.1794 \\

      &
      &  CHISEL-4
      & 1.93 M
      & 0.2370
      &  0.1630
      &  0.1401
      &  0.1279 \\

      &
      &  CHISEL-7
      & 5.79 M
      & 0.2183
      &  0.1138
      &  0.0957
      &  0.0878 \\

      &
      &\cellcolor[HTML]{EFEFEF} $\bigstar$ ConvNeXt-S
      &\cellcolor[HTML]{EFEFEF} \bf  1.26 M
      &\cellcolor[HTML]{EFEFEF}  0.2018
      &\cellcolor[HTML]{EFEFEF}   0.1077
      &\cellcolor[HTML]{EFEFEF}  \bf  0.0838
      &\cellcolor[HTML]{EFEFEF}  \bf  0.0768 \\

      &
      &\cellcolor[HTML]{EFEFEF}  $\bigstar$ ConvNeXt-L
      &\cellcolor[HTML]{EFEFEF}  4.98 M
      &\cellcolor[HTML]{EFEFEF}  0.2059
      &\cellcolor[HTML]{EFEFEF} \bf  0.0876
      &\cellcolor[HTML]{EFEFEF} \bf  0.0733
      &\cellcolor[HTML]{EFEFEF} \bf  0.0661 \\
      \hline
      \hline

      \multirow{2}{*}{LoS}
      &  \multirow{2}{*}{Non-stationary}
      &  CHISEL-7
      & 5.79 M
      &  0.1783
      &  0.1512
      &  0.1460
      &  0.1431 \\

      &
      & \cellcolor[HTML]{EFEFEF} $\bigstar$ ConvNeXt-L
      &\cellcolor[HTML]{EFEFEF}  4.98 M
      &\cellcolor[HTML]{EFEFEF}   0.1781
      & \cellcolor[HTML]{EFEFEF} \bf  0.1013
      & \cellcolor[HTML]{EFEFEF} \bf  0.0908
      & \cellcolor[HTML]{EFEFEF}\bf  0.0958 \\
      \hline
      \hline

      \multirow{2}{*}{\makecell{Mixed LoS\\and NLoS}}
      &  \multirow{2}{*}{Non-stationary}
      &  CHISEL-7
      & 5.79 M
      & 0.2643
      & 0.2018
      & 0.1941
      & 0.1960 \\

      &
      &  \cellcolor[HTML]{EFEFEF}{$\bigstar$ ConvNeXt-L}
      & \cellcolor[HTML]{EFEFEF} 4.98 M
      & \cellcolor[HTML]{EFEFEF}  0.1418
      & \cellcolor[HTML]{EFEFEF} \bf 0.0807
      & \cellcolor[HTML]{EFEFEF} \bf  0.0751
      & \cellcolor[HTML]{EFEFEF} \bf 0.0746 \\
      \hline
      \hline

    \end{tabular}
       \begin{tablenotes}[para,flushleft]
         \item {{In this table, CHISEL-1, CHISEL-4, and CHISEL-7 represents multiplying the number of channels in each convolutional layer of CHISEL in \cite{[49]} by $1$, $4$, and $7$, respectively. The settings for the ConvNeXt-S (small) and the ConvNeXt-L (large) are $C_1=C_2=C_4=3$, $C_3=27$, $E_1=128$, $E_2=256$, $E_3=512$, $E_4=1024$, and $C_1=C_2=C_4=3$, $C_3=27$, $E_1=256$, $E_2=512$, $E_3=1024$, $E_4=2048$, respectively.}}
   \end{tablenotes}
  \end{threeparttable}
  \end{center}
   \vspace{-0.6cm}
\end{table*}

\subsection{Extended to more practical scenarios}

In TABLE \ref{table}, we evaluate the performance of the proposed ConvNeXt-based scheme in practical mixed LoS and NLoS scenarios and compare it with the state-of-the-art DL-based localization technique CHISEL \cite{[49]}.
For modeling a mixed LoS and NLoS scenario, we assume the existence of one LoS path, one NLoS path, and one diffraction path.
The spatial non-stationary modeling at this time refers to equation 7 in \cite{[5]}.
It is worth noting that the inputs for the ConvNeXt and the CHISEL in TABLE \ref{table} involve $4$ beams sweeping for
angle estimation and $2$ beams sweeping for distance estimation. For simplicity, we have omitted the suffixes.
In addition, to compare the localization performance of ConvNeXt and CHISEL under the same complexity, we adjusted the number of convolutional channels in CHISEL.
Specifically, CHISEL-1, CHISEL-4, and CHISEL-7 represent scenarios when the number of channels in each convolutional layer of CHISEL \cite{[49]} is multiplied by $1$, $4$, and $7$, respectively.
The settings for ConvNeXt-S (small) and ConvNeXt-L (large) are $C_1=C_2=C_4=3$, $C_3=27$, $E_1=128$, $E_2=256$, $E_3=512$, $E_4=1024$, and $C_1=C_2=C_4=3$, $C_3=27$, $E_1=256$, $E_2=512$, $E_3=1024$, $E_4=2048$, respectively.

From TABLE \ref{table}, it is evident that the proposed ConvNeXt-L achieved optimal performance in three different scenarios. A comparative analysis reveals that spatial non-stationary characteristics reduce the localization performance of the ConvNeXt-L, whereas the presence of NLoS paths enhances its performance.
In contrast, CHISEL-7 does not achieve higher performance in mixed LoS and NLoS scenarios compared to LoS scenarios, likely due to its inability to learn deep-level features in complex environments.
In addition, we observe that compared to ConvNeXt-L in spatial non-stationary LoS scenario, ConvNeXt-S in the spatial stationary LoS scenario with only $25.30\%$ complexity of ConvNeXt-L achieves similar localization accuracy at low SNRs, but exceeded it at SNRs.
This highlights the significant impact of spatial non-stationary characteristics on localization performance.
More importantly, ConvNeXt-L can achieve higher localization performance with just $86\%$ of the computational complexity required by CHISEL-7.

\begin{figure}
  \centering
  \includegraphics[width=2.6in]{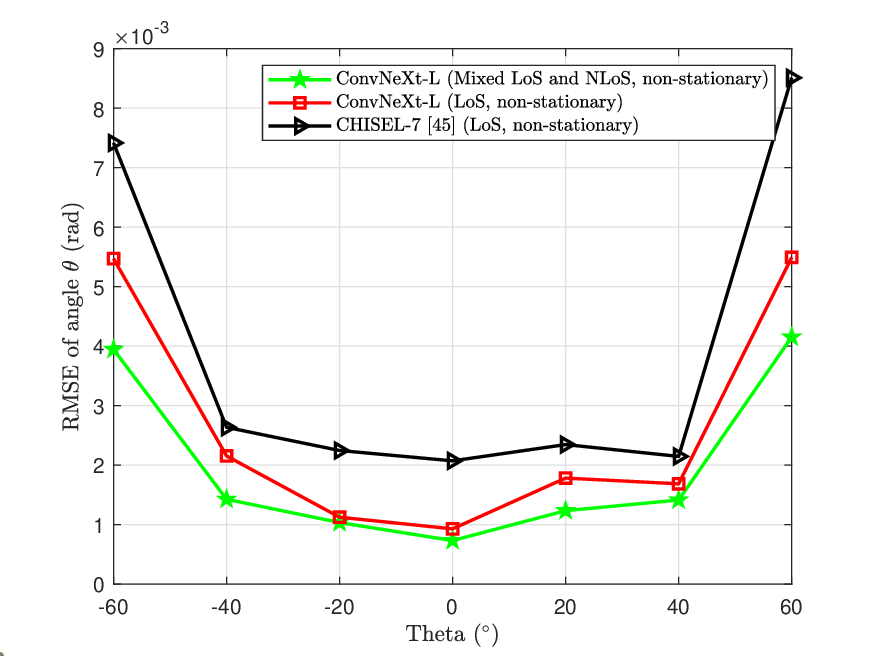}
  \caption{{{The angle RMSE performance comparison of the ConvNeXt method and CHISEL method against angles.}}}
  \label{angle}
\vspace{-0.8cm}
\end{figure}

In Fig. \ref{angle}, we evaluate the angle RMSE performance of the ConvNeXt and CHISEL methods across various angles. The distance of the UE is set at $ r = 15 $ m and the SNR is maintained at $ 10 $ dB.
A key finding is that regardless of the scheme or scenario, the RMSE of the angle decreases as $|\theta|$ decreases, with the minimum angle RMSE value occurring at the centerline of the ULA.
This improvement is attributed to the fact that the inputs for both the ConvNeXt and CHISEL schemes are derived from CBS-based beam sweeping techniques.
It is worth noting that, the CBS-based approach achieves angle grid search by directing beams from different subcarriers toward various desired angles.
More importantly, as shown in Fig. 4, the angle grid search features finer grid spacing near smaller values of $|\theta|$.
This characteristic enables the ConvNeXt and CHISEL schemes to achieve the minimum RMSE at $\theta = 0$.
In summary, our proposed ConvNeXt-based scheme achieves optimal localization performance across various scenarios when compared to other state-of-the-art localization technologies, such as CBS in \cite{[3]} and the CHISEL in \cite{[49]}.


\section{Conclusions}\label{Conclusions}

In this paper, we derived the CRBs for user localization in a wideband XL-MIMO system.
Based on the beam squint signal model, we evaluated the effects of spatial non-stationarity, the number of subcarriers, and bandwidth on the CRBs for both angle and distance estimation.
Additionally, we proposed a user localization scheme, the CBS-BT, which leverages the near-field controllable beam squint.
Moreover, we introduced a ConvNeXt based user localization scheme to enhance accuracy by fully extracting information from multiple subcarriers.
Furthermore, we designed an angle estimation module based on ConvNeXt to reveal the impact of angle estimation accuracy on the CBS-BT performance.
Numerical simulations validated the effectiveness of our proposed schemes, particularly highlighting the achievement of centimeter-level user localization with the ConvNeXt based scheme.

\section*{Appendix}

This appendix derive the explicit derivation of $ {\vartheta _{k,rr}} = \frac{{\partial {\bf{u}}_k^H}}{{\partial {r_k}}}\frac{{\partial {{\bf{u}}_k}}}{{\partial {r_k}}} $, $ {\vartheta _{k,r\theta }} = {\vartheta _{k,\theta r}} = \Re \left\{ {\frac{{\partial {\bf{u}}_k^H}}{{\partial {r_k}}}\frac{{\partial {{\bf{u}}_k}}}{{\partial {\theta _k}}}} \right\} $, and $ {\vartheta _{k,\theta \theta }} = \frac{{\partial {\bf{u}}_k^H}}{{\partial {\theta _k}}}\frac{{\partial {{\bf{u}}_k}}}{{\partial {\theta _k}}} $.
Note that the remaining task for the derivation is to compute the terms $ {\frac{{\partial {{\bf{u}}_k}}}{{\partial {r_k}}}} \in {\mathbb{C}^{M \times 1}} $ and $ {\frac{{\partial {{\bf{u}}_k}}}{{\partial {\theta _k}}}} \in {\mathbb{C}^{M \times 1}} $.

For the first term, we have
\begin{equation}
\begin{aligned}\label{a1}
{\left[ {\frac{{\partial {{\bf{u}}_k}}}{{\partial {r_k}}}} \right]_m} 
 &= \sum\limits_{n = 1}^N {\left( {\frac{{\partial {\beta _{k,m}}}}{{\sqrt N \partial {r_k}}}{e^{j{k_m}r_k^{\left( n \right)}}}{e^{ - j2\pi {\phi _n}}}{e^{ - j2\pi {{\tilde f}_m}{t_n}}}{b_{k,n}}} \right)} \\
 &+ \sum\limits_{n = 1}^N {\left( {\frac{{\partial {e^{j{k_m}r_k^{\left( n \right)}}}}}{{\partial {r_k}}}\frac{{{\beta _{k,m}}}}{{\sqrt N }}{e^{ - j2\pi {\phi _n}}}{e^{ - j2\pi {{\tilde f}_m}{t_n}}}{b_{k,n}}} \right)},
\end{aligned}
\end{equation}
where $ {{k_m}} = {{2\pi {f_m}} \mathord{\left/ {\vphantom {{2\pi {f_m}} c}} \right. \kern-\nulldelimiterspace} c} $ and $ {{b_{k,n}}} $ is the $n$-th element of the vector $ {{\bf{b}}\left( {{\Theta _k}} \right)} $ in \eqref{5}.
We first compute $ {\frac{{\partial {\beta _{k,m}}}}{{ \partial {r_k}}}} $. We have
\begin{equation}
\begin{aligned}\label{a2}
\frac{{\partial {\beta _{k,m}}}}{{\partial {r_k}}} = \frac{{\partial \frac{c}{{4\pi {f_m}{r_k}}}}}{{\partial {r_k}}}
 =  - \frac{c}{{4\pi {f_m}{{\left( {{r_k}} \right)}^2}}} =  - \frac{{{\beta _{k,m}}}}{{{r_k}}}.
\end{aligned}
\end{equation}
Then, the term $ {\frac{{\partial {e^{j{k_m}r_k^{\left( n \right)}}}}}{{\partial {r_k}}}} $ can be computed as
\begin{equation}
\begin{aligned}\label{a3}
\frac{{\partial {e^{j{k_m}r_k^{\left( n \right)}}}}}{{\partial {r_k}}} 
&= j{k_m}{e^{j{k_m}r_k^{\left( n \right)}}}\left( {1 - \frac{{{{\left( {{{n'_n}}} \right)}^2}{d^2}{{\cos }^2}{\theta _k}}}{{2{{\left( {{r_k}} \right)}^2}}}} \right).
\end{aligned}
\end{equation}
Substitution of \eqref{a2} and \eqref{a3} into \eqref{a1} yields
\begin{equation}
\begin{aligned}\label{a4}
{\left[ {\frac{{\partial {{\bf{u}}_k}}}{{\partial {r_k}}}} \right]_m} &=
\sum\limits_{n = 1}^N \Bigg[ \frac{{{\beta _{k,m}}}}{{\sqrt N }}{e^{j{k_m}r_k^{\left( n \right)}}}{e^{ - j2\pi {\phi _n}}}{e^{ - j2\pi {{\tilde f}_m}{t_n}}}{b_{k,n}} \\
& \quad \quad  \times \left( {j{k_m} - \frac{{j{k_m}{n^2}{d^2}{{\cos }^2}{\theta _k}}}{{2{{\left( {{r_k}} \right)}^2}}} - \frac{1}{{{r_k}}}} \right) \Bigg] \\
 & \buildrel \Delta \over = \xi _{k,m}.
\end{aligned}
\end{equation}
Then, we can obtain $ {\frac{{\partial {{\bf{u}}_k}}}{{\partial {r_k}}}} = {{\bm \xi} _{k}} = {\left[ {{\xi _{k,1}},{\xi _{k,2}}, \cdots ,{\xi _{k,M}}} \right]^T} $.

For the second term, we have
\begin{equation}
\begin{aligned}\label{a5}
{\left[ {\frac{{\partial {{\bf{u}}_k}}}{{\partial {\theta _k}}}} \right]_m} 
&= \sum\limits_{n = 1}^{N } {\left( {\frac{{\partial {e^{j{k_m}r_k^{\left( n \right)}}}}}{{\partial {\theta _k}}}\frac{{{\beta _{k,m}}}}{{\sqrt N }}{e^{ - j2\pi {\phi _n}}}{e^{ - j2\pi {{\tilde f}_m}{t_n}}}{b_{k,n}}} \right)} .
\end{aligned}
\end{equation}
Then, the term $ {\frac{{\partial {e^{j{k_m}r_k^{\left( n \right)}}}}}{{\partial {\theta _k}}}} $ can be computed as
\begin{equation}
\begin{aligned}\label{a6}
\frac{{\partial {e^{j{k_m}r_k^{\left( n \right)}}}}}{{\partial {\theta _k}}} 
&= j{k_m}{e^{j{k_m}r_k^{\left( n \right)}}}  \Big(  - {{n'}_n}d\cos {\theta _k} \\
& \quad \quad + \frac{{{{\left( {{{n'}_n}} \right)}^2}{d^2}}}{{2{r_k}}}\left( { - 2\sin {\theta _k}\cos {\theta _k}} \right) \Big).
\end{aligned}
\end{equation}
Plugging \eqref{a6} into \eqref{a5}, we obtain
\begin{equation}
\begin{aligned}\label{a7}
{\left[ {\frac{{\partial {{\bf{u}}_k}}}{{\partial {\theta _k}}}} \right]_m} \!\! & =  \sum\limits_{n = 1}^N \Bigg( \frac{{{\beta _{k,m}}}}{{\sqrt N }}{e^{j{k_m}r_k^{\left( n \right)}}}{e^{ - j2\pi {\phi _n}}}{e^{ - j2\pi {{\tilde f}_m}{t_n}}}{b_{k,n}}j{k_m}\\  &  \times \! \bigg( \!\!  - {{n'}_n}d\cos {\theta _k}
\! + \!\frac{{{{\left( {{{n'}_n}} \right)}^2}{d^2}}}{{2{r_k}}}\left( { - 2\sin {\theta _k}\cos {\theta _k}} \right) \bigg) \Bigg)  \\
& \buildrel \Delta \over = \zeta_{k,m}.
\end{aligned}
\end{equation}
Then, we can obtain $ {\frac{{\partial {{\bf{u}}_k}}}{{\partial {\theta _k}}}} = {{\bm \zeta} _k} = {\left[ {{\zeta _{k,1}},{\zeta _{k,2}} \cdots ,{\zeta _{k,M}}} \right]^T} $.

Finally, according to \eqref{a4} and \eqref{a7}, the terms $ {\vartheta _{k,rr}}  $, $ {\vartheta _{k,r\theta }}  $, and $ {\vartheta _{k,\theta \theta }}  $ can be computed as
\begin{equation}
\begin{aligned}\label{a8}
{\vartheta _{k,rr}} = \frac{{\partial {\bf{u}}_k^H}}{{\partial {r_k}}}\frac{{\partial {{\bf{u}}_k}}}{{\partial {r_k}}} = {\bm\xi} _k^H{{\bm\xi} _k},
\end{aligned}
\end{equation}
\begin{equation}
\begin{aligned}\label{a9}
 {\vartheta _{k,r\theta }} = {\vartheta _{k,\theta r}} = \Re \left\{ {\frac{{\partial {\bf{u}}_k^H}}{{\partial {r_k}}}\frac{{\partial {{\bf{u}}_k}}}{{\partial {\theta _k}}}} \right\} = \Re \left\{ {\bm\xi} _k^H {{\bm \zeta} _k}  \right\},
\end{aligned}
\end{equation}
\begin{equation}
\begin{aligned}\label{a10}
{\vartheta _{k,\theta \theta }} = \frac{{\partial {\bf{u}}_k^H}}{{\partial {\theta _k}}}\frac{{\partial {{\bf{u}}_k}}}{{\partial {\theta _k}}}= {{\bm \zeta} _k^H}{{\bm \zeta} _k},
\end{aligned}
\end{equation}
respectively.

\vspace{-0.6cm}
\bibliographystyle{IEEEtran}
\bibliography{IEEEabrv,Ref}

\end{document}